# A Biography of Henri Poincaré – 2012 Centenary of the Death of Poincaré

Galina Weinstein

On January 4, 2012, the centenary of Henri Poincaré's death, a colloquium was held in Nancy, France the subject of which was "Vers une biographie d'Henri Poincaré". Scholars discussed several approaches for writing a biography of Poincaré. In this paper I present a personal and scientific biographical sketch of Poincaré, which does not in any way reflect Poincaré's rich personality and immense activity in science: When Poincaré traveled to parts of Europe, Africa and America, his companions noticed that he knew well everything from statistics to history and curious customs and habits of peoples. He was almost teaching everything in science. He was so encyclopedic that he dealt with the outstanding questions in the different branches of physics and mathematics; he had altered whole fields of science such as non-Euclidean geometry, Arithmetic, celestial mechanics, thermodynamics and kinetic theory, optics, electrodynamics, Maxwell's theory, and other topics from the forefront of Fin de Siècle physical science. It is interesting to note that as opposed to the prosperity of biographies and secondary papers studying the life and scientific contributions of Albert Einstein, one finds much less biographies and secondary sources discussing Poincaré's life and work. As opposed to Einstein, Poincaré was not a cultural icon. Beginning in 1920 Einstein became a myth and a world famous figure. Although Poincaré was so brilliant in mathematics, he mainly remained a famous mathematician within the professional circle of scientists. He published more papers than Einstein, performed research in many more branches of physics and mathematics, received more prizes on his studies, and was a member of more academies in the whole world. Despite this tremendous yield, Poincaré did not win the Nobel Prize.

## 1. No Properly Written Biography of Poincaré

Jules Henri Poincaré was probably the most well-known nineteenth century French mathematician, physicist and philosopher. Yet there is no full-length biography of Poincaré, a biography which describes in a through manner his life, his scientific achievements and his other contributions to world-culture. All that one can find is a few old biographies of Poincaré in French, old biographical sketches written in memoriam of Poincaré after he had died prematurely at the age of 58 in 1912, and a few short and very succinct modern biographical sketches. These latter are often based on Poincaré's correspondence with his colleagues, obituaries to the press from 1912 describing his scientific and philosophical achievements, student notebooks from his student days, and other primary sources that are found in the Poincaré archives in Nancy, France and elsewhere. Old biographies were written by mathematicians or colleagues who knew Poincaré personally or met him in conferences.

Poincaré published general writings such as *Science and Hypothesis*, which were comprehensible by the general public, and were indeed read by as many people as possible in the world, and brought him much fame, his general books still did not bring him enough publicity to be a cultural world icon. He did not become a world hero and authors outside France did not bother to write comprehensive biographies of Poincaré.

**2. Old Biographies of Poincaré**

The first English biography of Poincaré's youth was published four months after Poincaré's death in November 1912 in the *Journal of the Royal Astronomical Society of Canada* by **Alfred Tennyson Delury**.[1] Delury was born in 1864 and died in 1951, a professor of mathematics at the University of Toronto, and he probably met Poincaré in one of the mathematics conferences, or else met one of his acquaintances.

Delury reproduced almost word for word the biographical sketch found in **Ernst Lebon**'s biography. On Poincaré's reception day to the Academy of Sciences on January 28, 1909, **Frédéric Masson**, the director of the French Academy, delivered a response speech to Poincaré's speech at the Academy. Masson asked Poincaré's classmates, relatives, and acquaintances to tell him stories about Poincaré, and they all gave him a picture of his life, "qui, en l'absence d'une biographie exacte, aura du moins l'avantage de la priorité".[2] Masson gave his speech to the Academy *while Poincaré was listening*. Masson said, "Vous êtes né, il n'y a guère plus d'un demi-siècle", "Vous sortez d'une race ancienne longtemps établie à Neufchâteau et depuis un siècle à Nancy". That is, "You were born…","You came from an ancient race".[3] Ernst Lebon's book contains an extract of this speech.[4]

It is evident from the above text that Masson had probably spoken with the psychiatrist, **Dr. Eduard Toulouse**, who published a book, *Henri Poincaré*, later that year in December 1909. Toulouse analyzes Poincaré's personality, and the first chapter of the book describes Poincaré's early youth.[5] This chapter is based on Toulouse's 1897 interviews with Poincaré. One finds similar stories having the same characteristics in Masson's speech and in Toulouse's book. Hence, it is reasonable to assume that when Masson interviewed Poincaré's acquaintances one of them was Dr. Toulouse.

Toulouse's examination of Poincaré was performed in 1897. However, Toulouse published his book only in 1909.[6] Toulouse explained that for many years he hesitated to publish the notes. He said that he decided finally to publish the examinations after he had found things he had not seen before:[7] This was after Poincaré had published his 1908 paper dealing with the process of his invention, "L'invention mathématique", quoted in Toulouse's book on pages 183-186.[8] Toulouse realized that his own analysis from 1897 of Poincaré's process of creativity, which reflected the way he had arrived at his new ideas and created scientific papers, fitted

Poincaré's own description from 1908. He thus reexamined his notes and wrote the above said conclusions that appear on pages 186-187. Toulouse then said that he could organize his notes and publish the book.[A]

**Jacques Hadamard**, wrote in the introduction to his 1945 book, *The Mathematical Mind*, in which he also analyzed Poincaré's above 1908 paper "L'invention mathématique",[9]

"This study, like everything which could be written on mathematical invention, was first inspired by Henri Poincaré's famous lecture before the Société de Psychologie in Paris".[10]

Both Hadamard and Toulouse analyzed Poincaré's 1908 lecture from knowing personally Poincaré: Hadamard's acquaintance with Poincaré was collegiate, both were mathematicians; and Toulouse the psychiatrist interviewed Poincaré, while the latter told him many personal details on himself.

Another source is **Vito Volterra**, a 19th century mathematician, edited a collection of papers in 1914 and ended it with Poincaré's curriculum Vita. Volterra was a colleague of Poincaré. The contributors to this volume were: Volterra who discussed Poincaré's mathematical work, Jacques Hadamard wrote about Poincaré's research pertaining to the three body problem, **Paul Langevin** contributed a chapter on Poincaré the physician, and Poincaré's brother in law **Emile Boutroux** wrote a chapter about Poincaré's philosophical works.[11]

One of the oldest French biographies of Poincaré was written in 1925 by **Paul Appell**; this biography is based on an older sketch of Appell from 1912, and published later in 1921.[12] Appell was a professor of rational mechanics at the Sorbonne. Between 1920 and 1925 he was the president of the Academy of Sciences. He was born in Strasbourg, but when Germany annexed in 1870 half of Loraine, Paul Appell was sent to Nancy. There he met Poincaré and there began a long friendship between the two. Appell met Poincaré when they studied for the special entrance exams in mathematics

---

[A] (Footnotes only compare between Poincaré and Einstein). In 1916 when the Gestalt psychologist Max Wertheimer interviewed Einstein, in much the same way as Toulouse interviewed Poincaré. Like Toulouse, Wertheimer hesitated to publish the interviews with Einstein and his analysis of Einstein's creativity. It took Wertheimer a long time to publish the interviews with Einstein: Wertheimer died on December 10, 1943, and Wertheimer's book, *Productive Thinking*, was published posthumously in 1945. Wertheimer, Max, *Productive Thinking*, 1916/1945, New-York: Harper & Brothers. The chapter on Einstein, Chapter VII, "Einstein: The Thinking That Led to the Theory of Relativity", provides a report based on interviews made in 1916. Wertheimer published the book and Chapter VII only after Einstein commented on an early draft of this chapter. Norton, John, "Einstein's investigations of Galilean covariant Electrodynamics prior to 1905", *Archive for the History of Exact Sciences* 59, 2004, pp. 45-105; p. 77.

for the Grandes Écoles – the most prestigious technical university institutions in France.[13]

Another quite popular biography (in English) was written by **Tobias Dantzig**, *Henri Poincaré Critic of Crisis Reflections on his Universe of Discourse*.[14] Dantzig wrote the book in 1954 on the one hundredth anniversary of Poincaré's birth. He knew Poincaré as a professor from the Sorbonne. Dantzig was a student between 1906 and 1910 at the Sorbonne and he attended Poincaré's lectures. Like previous biographers of Poincaré, Dantzig reproduced some of the stories that are found in Lebon's (Masson's) and Toulouse's books. However, the reader gets an impression that Dantzig had also picked up on some gossip on Poincaré and heard some stories during his stay at the Sorbonne.

The above mentioned writers belonged to the old generation of writers. They were all mathematicians who met or knew Poincaré personally. The writers were *not* professional historians or professional writers.

An additional French biography, published in 1956, is **André Bellivier**'s biography.[15] This biography describes in great detail the young Poincaré. Bellivier already belonged to a generation of writers who supported their claims by documentary material from archives; he read primary sources, and this gave him valuable information about Poincaré's work and personal life, and assisted him in the evaluation of the archival documents.

However, Bellivier's biography does not deal with Poincaré's life beyond the year 1882. It therefore does not encompass Poincaré's immense work on the electrodynamics of moving bodies and the theory of the electron, and his writings pertaining to the principle of relativity.

Later sources have been written by historians and philosophers of science, and these discuss Poincaré's life and work. These sources are already mostly written in English, but many papers are still published in French. For instance, in 2000, the French edition of *Scientific American* dedicated a volume to Poincaré's life and science; one of the articles in this volume gives a summary of Poincaré's youth and adolescence.[16]

**Scott Walter** from the Poincaré Archives in Nancy and other scholars from the Poincaré Archives produced lately the collection of Poincaré's correspondence with other scientists. Walter also published many papers discussing Poincaré's science and life. **Arthur Miller** analyzed Dr. Toulouse's interviews with Poincaré in his book *Insights of genius*.[17] **Jean Mawhin** from Brussels also wrote a biographical sketch of Poincaré and mentioned the interviews with Toulouse.[18]

Another source is **Peter Galison**'s book, *Einstein's Clocks, Poincaré's Maps.* In his book Galison focuses on Poincaré's work in the Bureau des longitudes starting from 1893.[19] And this list by no means exhausts the literature about Poincaré's life; I only brought some examples.

### 3. Dreamy and Shy

Dr. Toulouse describes Poincaré in the following way,

"M. H. Poincaré is a man of size (1.65)

And body size (70 kilos with clothes)

Medium bent, slightly prominent belly.

Colored face, large red nose. The hair is brown and the mustache is blond.

[…]

The main measures are taken by

Dr Manouvrier. We can compare them with
other measures".

And then Toulouse brings a table of measures.[20]

Toulouse says[21] that Poincaré was not a great orator,[B]

"Mr. H. Poincaré speaks correctly, but with a shyness which he is aware of.
So he avoids speaking unprepared in public, except in circles of scientists. When he has to give a speech, as he has done at Jubilee Hermite, he writes in advance. At least he prepares a number of sentences and he pronounces and says them aloud to himself, usually at the beginning of his speech. When he talks the speech, it goes easily but he never had the idea to write down the whole speech and to learn it by heart. He primarily seeks clarity before correctness, and he thus repeats the words so that it will be clearer".

Toulouse continues to describe Poincaré,[22]

---

[B] In 1954 Einstein wrote, "Also, I never exactly became an orator later," but, Einstein wrote in the same letter: "My parents were worried because I started to talk comparatively late, and they consulted the doctor because of it. I cannot tell how old I was at that time, but certainly not younger than three". Hoffmann, Banesh and Dukas, Helen, *Albert Einstein Creator & Rebel*, 1973, New York: A Plume Book, 1973, p. 14. However, as opposed to Poincaré, Einstein "always liked to improvise", said his son Hans Albert, "for instance, when he had to give a talk he never knew ahead of time exactly what he was going to say. It would depend on the impression he got from the audience in which way he would express himself and into how much detail he would go". Whitrow, G. J. (ed), *Einsetin: the man and his achievement. The BBC Third programme Talks*, 1967, London: British Broadcasting Corporation, Forward by Christopher Sykes, p. 19.

"In practical life he shows discipline, and this is an essential element of his character. In terms of ideas and decisions he is hesitant. He is making his mind quickly, otherwise if he hesitates in the meantime it will be difficult for him to decide.

He is not ordered, although he appreciates the value of this quality".

In addition, "He loves music and his favorite composer is Wagner. In his youth he played a little piano. He draws a little".[23]

Tobias Dantzig wrote,[24]

"I saw him often between the years 1906 and 1910, when I was a student at the Sorbonne. I recall above all his unusual eyes: myopic, yet luminous and penetrating. Otherwise, my memory is that of a man small in stature, stooped and ill at ease, as it were, in limb and joint. This last impression was accentuated by his manner of writing on the blackboard. For, his penmanship was very bad, and his draftsmanship even worse. He was ambidextrous, and I recall an ironical remark of a fellow-student to the effect that Poincaré could use either hand with equal ease and dexterity".

Toulouse asked Poincaré to draw and sketch all sort of things – geometric figures, faces, figures, and so on; and he said that Poincaré is not talented in this direction, and indeed the drawings in Toulouse's book disclose Poincaré's lack of drawing talent. Toulouse then said,[25]

"He does not have the gift of mining".

This is curious because Poincaré chose to study at the École des Mines after he graduated and arrived the second out of hundreds at the École Polytechnique.

Finally according to Toulouse,[26]

"He makes daily interruptions for 2 to 3 hours. Also Poincaré sleeps every night from 10 at night to 7:00 in the morning. He tries to get 7 hours of sleep".

**4. A Mathematics Monster**

Poincaré was born on April 29, 1854 in Nancy, a town in the Lorraine County in France. He was born in the Hôtel Martigny, an apartment building, which still exists at the corner of Grande-Rue and rue de Guise. Today this building has transformed into a drugstore.[27]

Masson told that Poincaré used to say that his family name was in fact Pontcaré.[28] Poincaré's family was well-known in Lorraine. His grandfather, on the father side, Jacques-Nicolas was a pharmacist, his father, Léon, a neurologist, a professor at the Faculty of Medicine, and his uncle Antoni graduated l'École Polytechnique. He was an inspector of roads and bridges, and he had two sons, Lucian and Raymond. The first was to be a physicist and then a rector of the University of Paris. Raymond, was a prime minister and minister of foreign affairs and then the president of French Republic between 1913 and 1920. In 1895 Henri Poincaré's younger sister, Aline, married the famous moral philosopher Emile Boutroux, with whom Poincaré used to discuss philosophical problems. Their son Pierre was a mathematician and philosopher. Henri's mother, Eugénie Launois, came from a family of gentlemen farmers in Arrancy. She was twenty four years when Henri was born, and he resembled her in several physical characteristics.[29]

Toulouse wrote that several males in the family were distinguished.[30] Indeed the father was a professor at the faculty of medicine; the cousin was Raymond Poincaré, a statesman. His sister's son was a gifted mathematician. When performing the interviews with Poincaré – in 1897 – women could not freely study in universities and they did not have equal opportunities.

Poincaré had only one sister, a few years younger than him, and she was very intelligent.[31]

It was soon realized that Poincaré is a prodigy, and started to talk at nine months.[32]

Masson probably heard from one of Poincaré's relatives the following story, and then Delury appeared to have translated it to English in his sketch:[33] the story runs that when only nine months old, one evening, as darkness fell, his eye caught a star and he called his mother's attention to it; as others shone out he espied them and became excitedly intent in his search for others, so much so that he was with difficulty quieted, and brought into a mood for sleep. In this, one may choose to see a presage, so writes Delury.

On the basis of Masson's report (Lebon's biography),[34] Delury then said that later at the age of five a serious illness threatened for a time to deprive him of the power of speech; it left him weak, and intensified in him a certain native shyness. Illness often has a maturing influence, and it seems to have affected the child Poincaré in this way.[35] At the age of five he had a serious diphtheria paralysis of the soft palate and paraplegia which lasted for three months. For two months as a result of this paralysis, he presented trouble speaking.[36] Two years later after infection, he felt dizziness to go downstairs.[37]

At six, he could read and probably write, he believes he learned to remember easily counting.[38]

Masson also said (and Delury appears to have translated the latter report to English), "I was told" that Poincaré was a tender and shy child, he preferred the company of his sister upon playing with other children, and he was never drawn to violent sports. Toulouse reported that Poincaré used to invent games for his sister and his cousins. According to Toulouse, a private teacher gave Poincaré early mathematical skills. He immediately was interested. The private teacher taught him a few subjects. Masson also mentioned this teacher:[39] Poincaré received an emeritus teacher who came home, and was a friend of the family. Poincaré was asked by his teacher to do written assignments, and he had conversations with him, talking about everything. This teaching was so encyclopedic that it was appropriate to Poincaré's nature.[C]

Toulouse wrote that Poincaré came well-prepared on October 1862 to the Lycée in Nancy. He stayed there eight and half years and finished ninth.[40] Poincaré entered the Lycée of his native city Nancy at the age of nine, and followed his studies there for eleven years until 1873 (age 17). The Lycée was later renamed the Lycée Henri Poincaré.[41] Masson reported[42] that there Poincaré was one of the top students and was superior to his fellow students in all branches, exhibiting unusual ability in mathematics.[D]

Recall that Toulouse said on Poincaré, "In practical life he shows discipline, and this is an essential element of his character".[43] Poincaré seemed to show discipline and he was thus successful. However, in writing and drawing he was not so successful. He, however, easily learned his lessons; but perhaps he forgot quickly. He learned regularly for competitions, and he learned faster than other students.[44]

---

[C] Einstein the child also preferred to play with his sister at home and refrained from joining the boisterous games of children of family and relatives; he occupied himself with quieter things, and he "would play by himself for hours". Winteler-Einstein Maja, *Albert Einstein –Beitrag für sein Lebensbild*, 1924, reprinted in abridged form in *The Collected Papers of Albert Einstein Vol. 1: The Early Years,1879–1902* (*CPAE*, Vol. 1)*,* Stachel, John, Cassidy, David C., and Schulmann, Robert (eds.), Princeton: Princeton University Press, 1987, *CPAE*, Vol. 1, p. Xviii; Vol. 1, p.1vii. Einstein seemed not to be so tender like Poincaré. At the age of five or six Einstein also received his first instruction at home from a woman teacher. His sister Maja told the known story of him once grabbing a chair and stroking at his teacher, who was so frightened that she ran away terrified and was never seen again. Winteler-Einstein, *CPAE*, Vol. 1, pp xlviii-lxvi; Vol. 1, p. 1vii. Stories like these reflect Einstein's rebellious temper and lack of obedience.

[D] When Einstein was nine, he was not considered extremely talented. Frank, Philip, *Einstein: His Life and Times*, 1947, New York: Knopf, 2002, London: Jonathan, Cape, p. 10; Frank, Philip, *Albert Einstein sein Leben und seine Zeit*, 1949/1979, Braunschweig: F. Vieweg, p. 23. Of course Einstein was very likely more talented than the rest of the pupils in his class. Rather his teachers at school did not fully appreciate Einstein because of his lack of obedience and discipline. Reiser wrote about Einstein, "But as a student of languages, Albert was only mediocre. He lacked the phonetic, as well as the mnemonic, gift. He hated the burden of so much memorizing and did not show the slightest talent for learning by rote, which study of classical languages particularly called for". Reiser, Anton *Albert Einstein: A Biographical Portrait*, 1930/1952, New York: Dover p. 37. As opposed to Poincaré, Einstein was mediocre from the verbal point of view. Einstein refused to learn topics, of which he was weak and did not excel. Einstein's parents registered him to a Gymnasium, with its emphasis on classical languages and literature, and not to a Realschulen with an emphasis on natural sciences. Einstein did not like humanistic subjects and thus was considered not so talented in the Gymnasium.

9Masson said that,[45] Poincaré was best in Letters, and one of the teachers, a historian lured him for his studies. But when he read a whole treatise of geometry alone his teacher ran to his mother astonished and said, "Your son will be a mathematician".[E] Masson's description was greatly exaggerated by Delury saying that each year of Poincaré's work at the lycée strengthened the faith in this prediction.[46] Delury might have heard this and other stories as myths spread among mathematicians in conferences.

Toulouse wrote that early at school Poincaré was attracted to physics and natural sciences, but the gift as a mathematician began earlier.[47] And he began to read mathematical books intended for specialists. From childhood he showed ease to arithmetic.[48] He read early in school popular science books, and later more serious books. He took great pleasure at listening to music, he even learned to play the piano, but he was unable to play good.[F]

Poincaré won first prizes in the *concours général* (a competition between students from all French lycées). His professor at Nancy is said to have referred to him as a "monster of mathematics".[49]

---

[E] With Einstein his Latin teacher told him that nothing would turn out of him ("Einstein, aus Ihnen wird nie was Rechtes warden!"), and the other teachers were annoyed of him as well. Seelig Carl, *Albert Einstein: A documentary biography*, Translated to English by Mervyn Savill 1956, London: Staples Press, p. 12; Seelig Carl, *Albert Einstein; eine dokumentarische Biographie*, 1954, Zürich: Europa Verlag, p. 15. Indeed they were all annoyed because of his lack of obedience. On the other hand, *at first* only one professor thought that Einstein would become a great physicist. When Einstein recounted the entrance examinations for the Zurich Polytechnic that he took on October 1895, he said that the physicist Heinrich Friedrich Weber invited him to attend his college physics lectures, provided he remained in Zürich. However, quite soon Weber was disappointed of Einstein, also because of his rebellious character. And so Poincaré was initially very successful; but Einstein's defiance of authority explains why for him success did not come so easily. Einstein, Albert, "Erinnerungen-Souvenirs", *Schweizerische Hochschulzeitung* 28 (Sonderheft), 1955, pp. 145-148, pp. 151-153; Reprinted as, "Autobiographische Skizze" in Seelig Carl, *Helle Zeit – Dunkle Zeit. In memoriam Albert Einstein*, 1956, Zürich: Branschweig: Friedr. Vieweg Sohn/Europa, pp. 9-17; p. 9; Winteler-Einstein, 1924, *CPAE*, Vol. 1, p. xxii; *CPAE*, Vol. 1, p. 1xv.

[F] It is a well known fact that Einstein was a lover of music (violin playing); he was also fascinated with popular science books from early age. Max Talmud recalled, "He showed a particular inclination towards physics and took pleasure in talking about physical phenomena. I gave him therefore as reading matter, A. Bernstein's *Popular Books on Physical Science* [...]". Subsequently, Einstein began to read mathematical books intended for specialists. Talmud says that "contrary to popular belief he [at the age of 13] had an unusual predilection for mathematics, and because of this fact I gave him, after his promotion to the fourth grade, Spieker's textbook on geometry". Talmay, Max, "Personal Recollections of Einstein's Boyhood and Youth", *Scripta Mathematica*, New York, 1932, vol 1, pp. 68-71; p. 69.

According to Toulouse Poincaré learned languages easily. He learned German in the lycée, but with family he did not speak German. He improved this language during the war and also during the three months stay he made in Austria in 1887.[50] Delury reported that it was while Poincaré was still in the lycée that he was a witness of a humiliation of his native land, Nancy itself being occupied by the victorious Germans. With some of the hideous realities of war Poincaré came into close contact while assisting his father (who was a physician) and was caring for the sick and wounded. Scanning the war bulleting and interpreting to himself the reports of the battle operations in the papers that were within his reach, he came to understand the German language. Nancy was in the territory occupied by the Germans after the war; but later it was announced the withdrawal of foreign troops.[51] Indeed this was first described by Masson. Poincaré was sixteen, and the war broke. Every day he accompanied his father in his ambulance and he was his secretary. Poincaré who was interested in the news, read newspapers and this way he learned German.[52]

Dantzig also reproduced the same story, which he probably read in Delury's sketch; but Dantzig added the following anecdote,[53]

"Yes, there was much grist for rancor to feed on. So, let us look at the record. During the occupation Poincaré taught himself to read German. In fact, he learned the language so well that he could read German newspapers fluently: in this way he managed to stay abreast of events and to brief his friends during the period when French journals were banned. In the years to come he kept up his German studies, and it may be said without exaggeration that no Frenchman had a better knowledge of German mathematical and physical literature than Poincaré. And this takes me to the first episode of this queer story.

One of the earliest mathematical achievements of Poincaré was a generalization of the so-called *elliptic functions*. He could have called these functions ultra-elliptic, or pan-elliptic: he chose instead to call them *Fuchsian*, in honor of the German, Fuchs, who had indicated their possibility without proving their existence. This homage must have raised many a French eyebrow, as evidenced by an epigram which I heard twenty years later. This would read in free translation: "The only Fuchsian claim to fame is that a French discovery bears his name". These protests did not disturb Poincaré: for, a few months later he presented the French Academy of Sciences with a new discovery which he called *Kleinean Groups*, honoring the German mathematician, Felix Klein. There were several similar incidents where, it seemed, Poincaré had been leaning backwards to pay homage to German scientists: the most striking of all occurred at Göttingen in 1909, when Poincaré delivered a series of six lectures, *five of these in German*!"

However, the sixth lecture discussing the principle of relativity was delivered in French![54]

Toulouse reported that by the age of fifteen, Poincaré studied English. He went to England one day, and he tried to read signs using German words. At his return, he began to methodically learn English, following – later – a course of English at the School of Mines.[55]

In November 1871 Poincaré graduated school. He had taken the degrees of bachelor of Letters and Bachelor of sciences. Poincaré was eighteen. He was later asked by Dr Toulouse about his opinion on religion at that time.[56] Poincaré replied that he had first believed in religion and then he gradually came to doubt it. At the age of eighteen, he stopped believing in religion. At that stage he was a free thinker, he believed in the right to search and tell the truth, and for that reason he opposed to the clerical (that is, catholic) intolerance.[G]

After graduating the lycée, Poincaré entered a preparation class in mathematics for the concourse exams to the Grandes Écoles of Paris.

**5. Henri Poincaré and Paul Appell**

On December 22, 1912, a few months after Poincaré's death, his close friend Paul Appell sat down to write the following memorial sketch on their close friendship in Nancy during the preparation class for the concourse exams to the Grandes Écoles of Paris. In this sketch Appell tells in all sincerity his recollections, but his stories often sound like hero-worship to Poincaré, and as such are sometimes exaggerated. In fact, Poincaré was so brilliant in mathematics that, his friends exaggerated when telling stories about his mathematical abilities. And so Appell's report reveals that Poincaré's friends were immensely impressed with his exceptional mathematical abilities and knowledge. For instance, Appell said that Poincaré won the first prize in the elementary mathematics concourses competitions; he was the only one to solve the final year problem given by the École Polytechnique.[57]

Appell starts his report by describing the meeting with Poincaré,[58]

"After Easter, my mother sent me from Strasbourg to Nancy, to participate in the preparatory class for the École Polytechnique. I arrived, as an inexperienced young student to the class of Mr. Pruvost, who was willing to admit me willingly, even though the courses were very advanced, and he gave me advices for which I owe him a great recognition. At the beginning of October, the special class was assigned to one of the most distinguished young professor, Elliot, a mathematician, whose value had an effect on all students". And then, "From the very first class, one of my friends told me, pointing to Poincaré: "this is a very strong type, he will be accepted second to the École Forestière". Meaning strong in learning, because the school of Forestry was prestigious among the students of Nancy.[59] And Masson also reproduced the same anecdote.[60]

---

[G] One can discern common points between Poincaré's and Einstein's attitude to religion. When Einstein was twelve-years-old (despite his parents being secular and completely against matters of religion) a deep religious feeling was awakened in him after he received Jewish religious instruction by a distant relative. Winteler-Einstein, 1924, *CPAE*, Vol. 1, p. xx; *CPAE*, Vol. 1, p. 1x.  However, a year later, Einstein's attitude toward religion experienced an important change. He regarded ritual customs as superstitious usages, preventing man from thinking independently. There arose in Einstein an aversion to the orthodox practices of the Jewish or any other traditional religion, as well as to attendance at religious services, and this he never lost. Frank, 1947/2002, pp. 14-15; Frank, 1949/1979, pp. 28-29.

Appell then says: "The appearance of Poincaré struck me: At first glance, he was not the ordinary type of the intelligent student: he seemed to be absorbed in inner thoughts, with his eyes somewhat obscured by reflection: when he spoke, his eyes sparkled with an expression of kindness, which had both malicious and deep expression. I felt drawn to him as we were both outsider students; we exchanged a few words when going out the class. I was struck by his manner of speaking: short and a little choppy, and interspersed with long silences". It was obvious that, "From the first questions in class, his superiority appeared to brighten; he answered the questions by eliminating intermediate explanations, with short and concise answers, answers that the teacher always asked him to develop. The teacher used to tell him: "If you answer this way in the exam, you might be misunderstood". Appell then recalls, [61] "We used to talk, me and Poincaré, when leaving class, and soon we were completely attached.[H]

Appell describes their walks in Nancy's streets back from class to their homes with two additional friends, and their conversations on geometry and philosophy during their way back home. And then Appell says, "Poincaré had read much; he was studying J. Bertrand's algebra, Duhamel's analysis, Chasles' higher geometry, the geometry of Rouche. With great simplicity and loyal friendship he gave to his fellow students all the information and explanations they asked for. He had synthetic distinctions for problems, so when the professor requested the locus of points, where we see an ellipse at a given angle, Poincaré gave immediately [the answer]". [62]

---

[H] Einstein's superiority in class also brightened – the students (and *not* his experimental physics teacher Jean Pernet) immediately appreciated his superiority; because Einstein the rebel solved the problems by eliminating Pernet's explanations. Margarette von Uexküll, a fellow student of Einstein, studying biology, reported in 1956 that one day in the experimental course of Pernet, "She had spent the whole of a warm June afternoon wrestling with an experiment in the Polytechnic's laboratory. Frustration overwhelming her, she was drawn into an argument with a small, fat physics Professor [Pernet], who refused to let her seal a test-tube with a cork for fear it would break. Suddenly she noticed 'a pair of the most extraordinary large shining eyes that were clearly warning me'. These belonged to Einstein, who quietly assured her that the professor was mad and had recently fainted during an angry fit in front of his class. He suggested that she give him her laboratory notes so that he could cook up some better results. At the next review, the professor exclaimed. 'There, you see. With a little goodwill, and despite my impossible methods, you can apparently work out something useful' ". Highfield, and Carter, pp. 39-40; "*Erinnerungen von Margarette von Uexküll, Frankfurt Allgemeiner Zeitung*, 10 March 1956.

And "In problems of analytic geometry he often gave very elegant geometric solutions. Examples that come to my memory: The general concourses of special mathematics, Poincaré's composition was not only ranked first among the entire compositions in all departments in Paris, but was especially notified by the examiners". [63]

However, "When the exams approached, our teacher showed growing concern that Poincaré would give answers that were too elliptical, which might seem obscure to reviewers". [64]

Indeed Appell reports that, at the École Normale supérieure's concourse in Paris there was an amusing incident with Poincaré. Qualified applicants were required to sketch geometric descriptions during the oral examinations pertaining to their oral answers. [65] Poincaré, who did not see any interest in tracing mathematical lines and in making a careful design which bored him, preferred instead, after having all the data, to search for the equation of the horizontal projection of the intersection of curves. Hence he found the curves, with perfection not reached by those who had used the conventional constructions; but, when drawing it on his sheet, something distracted him and he put it in reverse by turning it around a $180^0$. The examiner was very intrigued by this solution, which was both inaccurate and perfect. [66]

Appell recalls that this examiner, "already dead today" (in 1912) thought that Poincaré expressed himself badly and that he would not be a good professor; "so he gave him such a low grade that to our amazement, he was ranked fifth at the École Normale supérieure; a strange fate for a genius who cannot meet the classifications of ordinary men!" [I] And then Appell mentions the story of the famous mathematician Galois, who was denied entrance to the École Polytechnique, because he entered into a polemic with his examiner on logarithms and then gave the right answer. [67]

"After the oral examinations at the École Normale, we returned between August 4 and 6, 1873 to Nancy to do written compositions for the École Polytechnique. We found the city in joy: flags everywhere, on all houses, vehicles, wagons of milk, and vegetable carts: German troops have left, and more specifically, while we went to do the compositions, the first troops of the French army have entered Nancy; a day of joy and deliverance, although sad for us the Alsatians, who could not forget that the liberation of French territory might not stop the Vosges for a long time". [68]

---

[I] Einstein's teacher again, Jean Pernet, gave Einstein the lowest possible grade, 1 in his course "Introduction to the practice of physics – elementary practice of physics". He once told Einstein, there was lack of capability in your work, he then told him that physics was difficult, and advised him to study medicine, law or philology instead. Seelig, 1956, pp. 40-41; Seelig, 1954, p. 47.

Appell recalled that, "Poincaré, who became nervous from excitement, especially managed poorly with this [written] composition, an exercise in which he did not excel. In addition, he had stuck his paper too fast, and then he stretched too rapidly the layers of papers with Chinese ink before the previous ones were dry. He was eager to join his family in the City Hall (Hôtel de Ville), there they awaited the arrival of French troops on the Place Stanislas".[69]

And then, "While we were preparing to the oral exams of the École Polytechnique, Poincaré, as a service, was willing to question his friends; he took the sheets of examinations and, while imitating the intonations of the examiners, and reproducing their habits of mind, he terribly stuck the sheets with glue, and then he laughed quietly". Subsequently he examined the candidates, gave them different oral problems to solve, and asked them to draw this and that. But then Appell recounts, "I can still see him saying, with tongue-in-cheek, to a candidate, a nearby colleague, terrified by this revelation, that the examiner Moutard asked about the properties of the rotation of the Limaçon of Pascal". [70]

Finally, "After very bright exams, and a particularly remarkable exam in geometry with Tissot, he was accepted at École Polytechnique first [among all othes]. We met again next fall in Paris, Poincaré was at the École Polytechnique, and I at the École Normale".[71]

**6. The École Polytechnique**

The École Polytechnique was created in 1794, and then called "École central des travaux publics" (Central School for Public Works). Since its foundation it was known for having great mathematicians among its teachers. For instance, in 1797 Joseph Louis Lagrange taught the "Theory of Analytical Functions", Jean Baptiste Joseph Fourier and August Louis Cauchy were also professors at the École Polytechnique.

When in 1873 the young Poincaré entered the prestigious École Polytechnique, its mathematical tradition echoed in all Europe. Nevertheless the most famous place to do mathematics was Göttingen, under Gauss' reign, and his students Dedekind and Riemann. Still the École Polytechnique was considered the best institute in France.[72]

The École Polytechnique produced a fraternity of alumni who were mathematicians and engineers that were called "Polytechnicians"; they received the best education in mathematical physics or mathematical technology. And after their graduation they joined the highest levels of the state's administrative structure.[73]

At that time, in 1874, the well-known philosopher Émile Boutroux arrived at the University of Nancy. In Nancy he met Poincaré's sister, Aline Poincaré, and they got married in 1875. Poincaré and Boutroux started to discuss philosophical problems, and the latter presented to the former philosophical writings as well as his friends, philosophers, and also colleagues.

Recall Paul Appell's stories; Poincaré was so brilliant in mathematics that people all around seemed to have exaggerated the stories of Poincaré's exceptional mathematical abilities. Toulouse wrote that "Poincaré so it seems took no notes at all in the courses of mathematics at the l'École Polytechnique". [74]

Masson also told the same story. Hence this story was probably known among Poincaré's students and colleagues: It is said that you followed your courses, at least in mathematics, without taking notes, without writing anything of what the professors were saying. [75]

Poincaré's science professors at the Polytechnique were Charles Hermite (Analysis), Marie-Alfred Cornu (Physics), Louis Jean Résal (Mechanics), Amédée Mannheim (Geometry), Hervé Faye (Astronomy) and Edmond Fremy (Chemistry). Poincaré's two most influential professors were Hermite and Cornu. [76]

Hermite was a professor at the École for quite a short time, between 1699 and 1876. He taught the course Analysis, a first year course at the École, the first course which Poincaré later gave when he started his academic career in 1879. Hermite also delivered the course differential geometry, and this topic became one of the favorite topics of Poincaré. In addition Hermite supervised Poincaré after he graduated. [77]

In lectures given by the young physicist Alfred Cornu, Poincaré was exposed to the fundamentals of the wave theory of light, including the problem of ether drag and the solution offered by Fresnel's drag coefficient. According to a retrospective account, [78] Poincaré designed and executed an optical experiment to demonstrate the translation of the Earth with respect to the luminiferous ether.[J]

In 1875 Poincaré graduated the École Polytechnique, had exceptionally high grades and was ranked second among hundreds of students. [79] He could easily finish first in class, but his poor drawing capabilities inhibited him and this is very likely the reason for why he was only ranked second and not the first. Scott Walter does not agree with this; he says that much has been made of Poincaré's putative inability to draw figures. Walter shows that numerous ink drawings and doodles in his school notebooks attest to Poincaré's ability to render with accuracy both solid objects and complex diagrams. The correspondence with his mother indicates that he made an effort to improve his drawing ability while at Polytechnique, and was satisfied with the result.[80]

---

[J] While a student at the Zurich polytechnic Einstein also performed an ether drift experiment. Einstein, Albert, "How I Created the Theory of Relativity, translation to English by Yoshimasha A. Ono, *Physics Today* 35, 1982, pp. 45-47; p. 46. According to Anton Reiser, Einstein "wanted to construct an apparatus to accurately measure the earth's movement against the ether". However, he was unable to do so because the skepticism of his teachers (Heinrich Friedrich Weber) was too great. Reiser, 1930, p. 52. It was thus probably a custom in those days among students to perform such experiments as part of the training in school laboratory.
A few years later, in the summer of 1881, Cornu met Albert Abraham Michelson in Paris. Cornu was expert in optical experimentation: trials made in France at measuring the velocity of light by Fizeau, Leon Foucault and others. Michelson met Cornu and Jules Jamin (from the Ploytechnique) and Éleuthère Mascart (professor of physics at the College de France) who performed first-order ether drift experiments, which yielded negative results.

Poincaré told Dr. Toulouse (as reported by the latter): He had ranked first when entering the École Polytechnique and then he graduated second.[81] Toulouse reported that in the réglementaires years, Poincaré started to work on mathematics, a major work he would publish ten years later under the title, *On the Theorem of M. Fuchs*.[82] We shall return to the Fuchsian function and Poincaré's exciting story of discovery of these functions.

**7. École des Mines**

During the second year of his studies Poincaré chose the future school after the Polytechnique, the École des Mines, the Mining School. The top-ranked graduates of Polytechnique almost invariably chose to continue their studies at this institution. There they studied three years towards a technical or administrative career in Mines. After graduating so high, Poincaré naturally decided to continue his studies at the École des Mines.[83] Recall that Toulouse wrote about Poincaré, "He does not have the gift of mining".[84] Nevertheless, Poincaré chose this direction and not the competing technical school École des Ponts et Chaussées; Antoine-Henri Becquerel graduated this latter school in 1894.

After three years of studies, graduates of the École des Mines were assured of high-profile careers offering excellent pay and social advantages, yet those who went on to become mine inspectors did so at great risk.[85]

One of Poincaré's professors at the Mines was Alfred Potier. Potier graduated the École Polytechnique like Poincaré, but a few years before him, in 1859. He then went straight to the École des Mines, again following the French tradition. However, unlike Poincaré, Potier was attracted to mineralogy, geology and mining engineering. He made numerous field studies in mining and until his retirement he was an inspector general. He remained attached to the geological survey.

Nevertheless, his favorite work, and the one he won reputation for was his work on mathematical and experimental physics. He studied the theory of heat, the theory of optics, the theory of light, and electricity. In this respect he had a lot in common with Poincaré. He considered the theoretical aspects of electricity, as well as the industrial aspects of the topic; but Poincaré was also interested in the technical aspects of electromagnetism when writing on Hertz's experiments with his antennas and spark gap detectors and receivers. Yet, even when Poincaré considered the applied aspects of electromagnetism, he did it as a mathematical physicist. On the other hand, Potier was more technical than Poincaré who was destined to be a mathematician. In 1893 Potier held the chair of industrial electricity at the École des Mines, while Poincaré would later hold in Paris the chair of mathematical physics. By 1881 Potier was also a professor of physics at the École Polytechnique, at the same time that Poincaré was a professor in Paris. Potier died in 1905.[86]

The study of mathematics was not part of the curriculum at the École des Mines. The notebooks of Poincaré from the period reflect a certain lack of concentration, and interest in the courses offered at the École des Mines. The notebooks are filled with hundreds of doodles.[87]

In addition, Poincaré studied geology and went to excursions. Masson said that Poincaré inherited from his father the passion for traveling. As a student at the École des Mines he went to missions in Austria and Sweden.

On one of these travels, one morning, Dr. Toulouse told this story:[88]

"He was often distracted: one day he put a sheet of white paper in an envelope instead of a letter; on another occasion he went on a trip and put in the trunk his bed sheet instead of a shirt. His family and apparently also other people knew that he was distracted".

Masson told the same story and probably heard it from Toulouse. Masson also borrowed Toulouse's analysis:[89] Toulouse generally described Poincaré as distracted.[K] Masson said[90] that when subsequently Poincaré traveled to parts of Europe, Africa and America, his companions noticed that he knew well everything from statistics to history and curious customs and habits of peoples.[L]

While studying at the school of Mines, Poincaré also got his diploma in mathematics from the Faculty of Science of Paris in August 1876. A year later in 1879, he graduated the École des Mines.[91]

**8. An Academic Career: Faculty of Science in Caen**

During the last two years at the École des Mines, Poincaré prepared his PhD thesis in mathematics, while studying on his own works of mathematicians.[92] He defended his thesis on August 1$^{rst}$, 1879 at the Faculty of Science in Paris in front of the jury: Pierre

---

[K] Einstein was not distracted. On the contrary, he had a tremendous ability to concentrate. When Einstein was 16 years old his sister reported that he already had this "remarkable power of concentration": even in a large, quite noisy group, he could withdraw to the sofa, take pen and paper, in hand, and lose himself so completely in a problem that the conversation of many voices stimulated rather than disturbed him. Winteler-Einstein, 1924, *CPAE*, Vol. 1, p. xxii; *CPAE*, Vol. 1, p. 1xiv.

[L] Einstein (as a grownup) did not appear to be an encyclopedic person like Poincaré: know everything in so many fields of research.

Bonnet, Jean-Claude Bouquet and Jean-Gaston Darboux. Darboux's report was very positive about the results and the methods, but was far less enthusiastic about the clarity of the style.[93] In 1879 Poincaré was awarded a doctorate in mathematics from the University of Paris. Poincaré's doctoral thesis was intended to improve a method for solving partial differential equations that had been suggested some years earlier by Cauchy.[94]

Poincaré's academic career started in the faculty of science in Caen Between August 1879 and October 1881. On April 20 1881, as he was about to leave, Poincaré married Louise Poullain d'Andecy. They had three daughters, Jeanne, Yvonne, Henriette, and then one son Léon.

Poincaré started to exhibit his talent for writing papers one after the other. He sent more than twenty notes to the *Comptes Rendus de l'Académie des Sciences de Paris*, dealing with three completely different topics: the arithmetic of forms, the qualitative theory of differential equations, and Fuchsian functions.[95]

Poincaré was inspired by the work of his teacher from school, Charles Hermite, to work on the Fuchsian functions. Many years later, in 1908, Poincaré presented his famous talk, "L'invention mathématique", to the Psychological Society of Paris at the Institut Général Psychologique. In this lecture Poincaré disclosed his own way of scientific thinking, and how in 1881 he came to solve the problem with the Fuchsian functions; functions which are known today by the name automorphic functions. Poincaré thus revealed in this talk the process of his scientific creativity,[96]

"For fifteen days I tried to demonstrate that any function could exist, similar to what I have called Fuchsian Functions, but I was very ignorant, and every day I sat at my desk. I spent an hour or two, I tried many combinations and I came to no results. One evening, contrary to my habit I took black coffee, I could not sleep; ideas rose in crowds; I felt them come against me, until two of them were clinging to form a stable combination. In the morning I had established the existence of a class of Fuchsian Functions, those derived from the hyper-geometric series; I had only to write the results, which took me a few hours.

Subsequently I wanted to represent these functions by the quotient of two series, the idea was perfectly conscious and deliberate; the analogy with elliptic functions guided me. I wondered what could be the properties of these series if they existed and arrived without difficulty at forming the series which I called Theta-Fuchsians.

At this time, I left Caen, where I was then Living, to take part in a geologic excursion conducted by the École des Mines. The events of the trip made me forget my mathematical work; having arrived at Coutances, we entered an omnibus to go for some sort of promenade; at the moment when I put my foot on the step, the idea came to me, without anything in my former thoughts seeming to have prepared for it, that the transformations I had used to define the Fuchsian functions were identical with those of non-Euclidean geometry. I did not verify this; I should not have had time, since, upon taking my seat in the omnibus, I resumed the conversation already started, but I felt a perfect certainty. On my return to Caen, I verified the result at my leisure to satisfy my conscience.

Then I turned to the study of arithmetic questions, apparently without much result, and without a suspicion that it might have anything to do with my previous research. Disgusted with my failure, I went to spend a few days at the seaside, and I thought of something else. One day, while walking on the cliff, the idea came to me, with just the same characteristics of brevity, suddenness and immediate certainty, that the undefined arithmetic transformations of quadratic forms were identical with those of non-Euclidean geometry".

Jaques Hadamard wrote the book, *The Mathematician's Mind*, and as mentioned in section 1, was inspired by Poincaré's lecture, "L'invention mathématique", from 1908. Hadamard translated the original French text to the above English one and commented,[97]

"Conditions of invention have been investigated by the greatest genius which our science has known during the last half century, by the man whose impulse is felt throughout contemporary mathematical science. I allude to the celebrated lecture of Henri Poincaré at the Société de Psychologie in Paris." Hadamard analyzed Poincaré's above text and said,[98]

"But that extraordinary fact of watching passively, as if from the outside, the evolution of subconscious ideas seems to be quite special to him. I have never experienced that marvelous sensation, nor have I ever heard of its happening to others".[M]

Dr. Toulouse remarked on the same text from Poincaré's 1908 talk: "This method of work is not common in science matters, and constitutes a very special character of mental activity of M. H Poincaré".[99]

Indeed Poincaré in his 1908 talk reflected on the way his unconscious arrives at his ideas.[100] "Most striking at first is this appearance of sudden illumination, a manifest sign of long, unconscious prior work. The role of this unconscious work in mathematical invention appears to me incontestable". Poincaré therefore does not sit near the table for too long. He takes a break and then returns, because most of the work is done by his unconscious; as he explained before, "but I was very ignorant, and every day I sat at my desk. I spent an hour or two, I tried many combinations and I came to no results".[101]

---

[M] Hadamard asked Einstein questions about his creativity process as well. Einstein told him that, words or language, as they are written or spoken, do not seem to play any role in his mechanism or thought. The psychical entities which seem to serve as elements in thought are certain signs and more or less clear images. The above elements are of visual and some of muscular type. Conventional words or other signs have to be sought only in a secondary stage. Hadamard, 1945, p. 140. Hence, both Einstein's and Poincaré's initial work is done by their unconscious. Einstein's first stage is *non-verbal*, and *visual*, while Poincaré's first stage is *verbal*, and includes mathematical signs and *non-visual* images (as we shall see below from Toulouse's explanation). Einstein's second stage is communicating his non-verbal images to sounding boards, while Poincaré's second stage is solitarily writing his results.

Dr. Toulouse analyzed the above trip and omnibus 1908 story of Poincaré solving the problem of the Fuchsian functions; and Toulouse explained that indeed Poincaré did not sit near the table for too long,[102]

"Mr. H. Poincaré works regularly morning from 10:00 to 12:00, and in the afternoon from 5:00 to 7:00, and never in the evening after dinner. He cannot work more than two fruitful consecutive hours. If he works more time it won't produce more. During the holidays he takes a complete intellectual rest. He also suffers from long sitting".

Toulouse compared Poincaré's own observations from his 1908 paper to his own interviews with Poincaré from 1897. He arrived at the following conclusions regarding Poincaré's creativity after quoting Poincaré's article on invention from 1908 on pages 183-186,[103]

"This is an uncommon way of working in scientific matters and it is a very special characteristic of the mental activity of M. H Poincaré. From all these facts, it appears that M. H. Poincaré's intellectual activity is especially spontaneous and automatic".

And, "Mr. H. Poincaré is not visual".[104] That is, according to Toulouse:[105] Poincaré did not have good *visual* memory. Nevertheless, Poincaré had a good memory, and he had a good *verbal* memory. Poincaré had a *logical* memory, in one story he did not invent facts.[N]

---

[N] In 1955 Einstein wrote, that as a pupil in Gymnasium he was neither particularly good nor bad. His principal weakness was a poor memory and especially a poor memory for words and texts. Hoffmann and Dukas, 1973, pp. 19-20. John Stachel argues that for Einstein, the process of thinking consists of two stages. The first stage "invention", primary non-verbal in nature; at a secondary stage, it was necessary for him to transform the results of this primary process into forms communicable to others. Stachel, John, *Einstein's Miraculous Year. Five Papers that Changed the Face of Physics*, 2005, Princeton and Oxford: Princeton University Press, Introduction, p. xxxv. According to Alexander Moszkowski Einstein told him, "For it is not true that this fundamental principle occurred to me as the primary thought. If this had been so perhaps it would be justifiable to call it a 'discovery'. But the suddenness with which you assume it to have occurred to me must be denied. Actually, I was lead to it by *steps* arising from the *individual* laws derived from experience". Moszkowski then says that Einstein supplemented this by emphasizing the conception "invention" and ascribed considerable importance to it: "Invention occurs here as a constructive act. This does not, therefore, constitute what is essentially original in the matter, but the

Toulouse described Poincaré's working habits: "there is no plan when he writes a paper";[106] and Toulouse explained,[107]

"The way he works, as a preparation he takes a few notes. Often there is no big plan. Or simply he writes a few ideas on a paper that he would develop. But usually he starts from memory, without his mind knowing the solution beforehand, before developing the problems that he would address. The beginning is usually easy. And he is then led by his work. At this time it is difficult to distract him".

Recall that Toulouse said that Poincaré was often distracted. Other people as well as family members were well aware that Poincaré was distracted.[108]

Dr. Toulouse wrote that when Poincaré searched for something, he often automatically wrote a formula to see if it stimulates ideas. If the start of the process was very painful, then Poincaré did not persist, and he abandoned the started work. That is why he had no patience. In some works of his, Poincaré proceeded by blows, taking and abandoning the subject. During the intervals he assumed that his unconscious continued the work of reflection. Stopping the work was difficult if there was not a strong enough distraction, especially if this work was not considered complete. For this reason, Poincaré did nothing important in the evening to avoid that his sleep be disturbed.[109]

Poincaré was thus aware of the way he could arrive at mathematical innovations, abruptly, with sudden strokes, during geological excursions and visits to mines or other trips, his great love from the days he used to study at the École des Mines. He did not need more than two hours of work per day, and most of the work was done unconsciously during these excursions and activities, and Poincaré could exploit his hobbies and his adventurous aspirations. His uniqueness was the ability to combine mathematics and adventurous aspirations within his creative unconscious mind.

**9. Professor at the Faculty of Sciences in Paris**

In the fall of 1881, Poincaré moved to Paris; there he stayed until his premature death. His career was boosted and his fame grew rapidly every year in France, and gradually

---

creation of a method of thought to arrive at a logically coherent system… the really valuable factor is intuition!". Moszkowski, Alexander, *Einstein the Searcher His Works Explained from Dialogues with Einstein*, 1921, translated by Henry L. Brose, London: Methuen & Go. LTD; appeared in 1970 as: *Conversations with Einstein*, London: Sidgwick & Jackson, 1970, p. 96; Moszkowski, Alexander, *Einstein, Einblicke in seine Gedankenwelt. Gemeinverständliche Betrachtungen über die Relativitätstheorie und ein neues Weltsystem. Entwickelt aus Gesprächen mit Einstein*, 1921, Hamburg: Hoffmann und Campe/ Berlin: F. Fontane & Co, p. 101. It seems that Toulouse exaggerated when he said that Poincaré did not invent facts ["Dans un récit, il n'invente pas de faits"], because like Einstein, Poincaré *invented* thought experiments. See his talk Poincaré, 1908a; and in *La Science et l'Hypothèse* and his other general books Poincaré proposed thought experiments profusely.

worldwide. He was named to the faculty of sciences in Paris as "Maître de Conférence d'Analyse" and in 1883 the duties of "Rérétiteur d'Analyse à l'École Polytechnique" were added. In 1885 he became "Chargé du Cours de Méchanique Physique et Expérimentale" at the Faculty of Sciences, and the next year, in 1886, he was appointed to the Professorship of mathematical physics and the theory of probability.

In the same year he was also appointed to professorship in celestial mechanics. As professor of mathematical physics, Poincaré, year after year, chose a new subject for his courses and dealt with the outstanding questions in the different branches of physics; he chose topics such as, hydrodynamics, thermodynamics and kinetic theory, optics, electrodynamics, Maxwell's theory, Hertz's experiments (and other topics from the forefront of Fin de Siècle physical science).[110]

Poincaré had taught electromagnetism at the Sorbonne, the École polytechnique and the École des postes et Télégraphes. Between 1889 and 1899 he thoroughly was teaching the works of Maxwell, Helmholtz, Hertz (the 1884-1890 interpretation of Hertz to Maxwell's equations), Larmor, Lorentz, and others in courses such as, *Électricité et optique* and *Théories électrodynamiques*. A favorite topic for Poincaré was Hertz's experiments with sparks; Hertz's detector and transmitter and propagation of electromagnetic waves, and the possibility of wireless telegraphy (Les théories de Helmholtz et les experiences de Hertz).[111]

Poincaré also published many papers on Hertz's experiments, discoveries and papers pertaining to electromagnetic waves, and Hertzian waves.[112]

Poincaré had a long exchange of letters with Heinrich Hertz on the topic of radio waves. The first letter of Poincaré to Hertz is dated August 15, 1890. Poincaré was excited of Hertz's experiments. However, he found a calculation error in Hertz's paper published in the *Wiedemann's Annalen*. He wrote Hertz and explained to him the error.[113] Hertz immediately replied, "Monsieur et trés honoré collègue".[114] Hertz told Poincaré, "The error you have discovered is truly unpleasant". Hertz told Poincaré that Oliver Lodge proposed a competing experimental approach to his own method.[115]

Poincaré approached Hertz again two weeks later on September 11, 1890. Poincaré told Hertz that he included his experiments in the volume containing his lessons from 1888, and that this latter volume provided teaching standards.[116] This was considered a great honor, because Poincaré's volumes of lessons were indeed the authorized textbooks in France, and they determined the teaching standards in mathematical physics. The lectures were mostly written by Poincaré's students, and afterwards he corrected and edited them; subsequently, these lectures were published in volumes, and he updated them according to new studies and innovations; for instance, new information that Hertz told Poincaré in their exchange of letters.

Poincaré also asked Hertz some questions. Hertz immediately replied to Poincaré on September 22, 1890. Hertz understood Poincaré's questions and actions as an important support for the views of Faraday and Maxwell.[117] Hertz ended his letter by writing, "please forgive me if my expressions are not always well chosen".[118] Hertz was writing his letters in French and not in German, and he thought he might have not written in proper French.

Two weeks later Poincaré replied to Hertz. Poincaré asked additional questions pertaining to Hertz's experiments. He wanted to include them in the volume of lessons that he was now publishing; Poincaré was publishing papers on Hertz's experiments, and he was thinking about the theory of telegraphy, which implemented Hertz's equipment. Poincaré therefore wrote Hertz, "voulez-vous me permettre maintenant quelques questions encore?", and asked Hertz additional technical questions on Hertz's experiments.[119] Hertz sent Poincaré a very long letter answering all his questions. In this letter he explained with sketches the procedure, difficulties at the moment of writing the letter, the errors that he did not know their causes, and the subtleties of his experiments.[120] Poincaré and Hertz continued to exchange letters on Hertz's experiments until the end of December 1891. On December 30, 1891 Hertz sent Poincaré the final letter.[121]

Louis De Broglie recalled in 1951 that all the young people in Poincaré's generation who were interested in mathematical physics (he remembered that there were little), were fed by Poincaré's textbooks. The teaching of mathematical physics at the Sorbonne was also based on Poincaré's books. Paul Langevin did not publish his courses.[122]

Poincaré specialized in mathematical physics. De Broglie distinguished between mathematical physics and theoretical physics: The first, according to De Broglie, is the profound and critical examination of the physical theories put forward by the researcher who assesses mathematical speculations in order to improve these theories and in order to render their inherent proofs more rigorous. In contrast, theoretical physics is the construction of theories suitable to serve as an explanation of the experimental facts and to guide the work of the laboratory staff. Extensive mathematical knowledge is a pre-requisite, although it is not, ordinarily, the work of real mathematicians; it requires wide knowledge of the experimental facts, and mainly some kind of intuition in physics, which not all mathematicians have, as did Poincaré. Poincaré, according to De Broglie, was especially destined to engage fruitfully in mathematical physics.[o]

In 1884 Poincaré first filed a request to be a member of the prestigious Academy of Sciences in Paris. However, only in 1887 he became a member of this Academy, and in 1906 he became its president; in 1908 the Académie Francaise opened its doors to him; Poincaré became its president until his premature death in 1912. Eventually he was a member of all the principal scientific societies and academies in Europe and the United States, and was awarded many prizes and medals by scientific bodies. [123]

---

[o] Einstein was a theoretical physicist according to De Broglie. De Broglie, 1951, p. 46.

## 10. The Bureau des Longitudes

Beginning in January 1893 until his death in 1912, Poincaré served at the Bureau des Longitudes in Paris.[124] It was a natural continuation to his adventurous love for geological excursions and navigations, which he had done during the time he studied at the École des Mines. In September 1899, about a year and a half after he published his well-known paper, "The Measurement of Time", Poincaré was elected president of the Bureau, a post he rose to again in both 1909 and 1910.[125]

One of the main projects of the Bureau des Longitudes was mapping places of areas in maps. Peter Galison reported[126] about the Bureau's publications on these tasks of determining the exact positions of colonies in 1897, coming into Poincaré's hands just before he wrote the 1898 paper, "The Measurement of Time".[P]

---

[P] Another bothering question is the following, could we compare Einstein sitting in the "Patentamt" in Bern and Poincaré sitting in the "Bureau des longitudes" in Paris (since 1894)? Einstein published his paper "On the Electrodynamics of Moving Bodies" in 1905 [Einstein, Albert, "Zur Elektrodynamik bewegter Körper, *Annalen der Physik* 17, 1, 1905, pp. 891-921] while sitting in the Patent Office; and Poincaré published his paper in 1898, "the measurement of time" [Poincaré, Henri, "La mesure du temps, *Revue de métaphysique et de morale* 6, 1898, pp. 371-384] while sitting in the Bureau of Longitudes. In fact one cannot do this comparison: neither compare between Einstein and Poincaré or between the two papers. On closer inspection, there is no similarity between Einstein sitting in the Patent Office and Poincaré sitting in the Bureau of Longitude (in addition to being in the academic world). Galison compares Poincaré and his work at the Bureau as a manager, and Einstein as a clerk at the Patent Office. Galison is nevertheless aware of the fact that, "In June of 1905, the contrast between Einstein and Poincaré could not have been greater. Poincaré was an Academician in Paris, fifty-one years old and at the height of his powers. He had been professor at France's most illustrious institutions, run interministerial commissions, and published a shelf of books – volumes on celestial mechanics, electricity and magnetism, wireless telegraphy, and thermodynamics. With over 200 technical articles to his name, he had altered whole fields of science. His bestselling volume of philosophical essays had brought his abstract reflections on the meaning of science to a huge audience, including Einstein. Einstein, at twenty-six, was by contrast an unknown patent officer, living in a walk-up flat in a modest section of Bern". Galison, 2003, p. 175. As to comparing between the two works: Martínez says that Einstein might have been inspired by patents of clocks, trains and clock towers that happened to stand near the Patent Office he used to work at in Bern, as alluded in Galison's book; but unlike Poincaré, Einstein himself left no written statement even implying that he became increasingly interested in time because of any timing technologies at the Patent Office. Martínez, Alberto, "Material History and Imaginary Clocks: Poincaré, Einstein, and Galison on Simultaneity", *Physics in Perspective* 6, pp. 224-

Galison in his book *Einstein's Clock's Poincaré's Maps*, connected among Poincaré's 1898 above paper, his work at the Bureau, and the technology of telegraphic signals. Poincaré *himself* developed a complete theory of télégraphie sans fil, especially from 1907, and in 1909, Poincaré explicitly did this connection in the *La Mécanique Nouvelle*.[127]

télégraphie sans-fil was a topic that much occupied Poincaré. In 1907 he wrote a book, *La théorie de Maxwell et les oscillations hertziennes; La télégraphie sans fil*,[128] and in 1908 he gave a series of lectures on the topic. The issues of the November 28, December 5, December 12, December 19, and December 26, 1908 of the Journal *La Lumière Électrique* published the lectures as series of papers by Poincaré under the title, "Conférences sur la Télégraphie Sans Fil" (Conference on Wireless Telegraphy). In the "Éditorial" introducing the papers the editor wrote that, in the summer of 1908, Henri Poincaré gave a series of lectures in a conference at the École Supérieure de Postes et Télégraphes, in which he outlined a complete theory of wireless telegraphy.

His intention was to outline a mathematical theory, which could facilitate the understanding of these phenomena.[129] Poincaré laid down the electromagnetic and technical foundations of wireless telegraphy following Marconi's work. He discussed the machinery for sending and reception of the signals, types of antennas, and the problems encountered with antennas. In 1911 Poincaré published in the same journal another series of papers on T.S.F, "Sur Diverse Questions Relatives A la Télégraphie Sans Fil".[130]

In his 1907 book Poincaré compared wireless Hertzian telegraphy with optical telegraphy.[131] He said that, first there are differences in wavelengths. Optical telegraphy focuses the light by means of lenses and mirrors, while Hertzian waves no longer do that. "les experiences de Hertz en 1888 qui ont fait enterer la question dans une nouvelle phase". While the wavelength is longer the diffraction is more significant and so the ability to bypass obstacles is greater. The biggest obstacle is the earth being round. Ordinary light can neither bypass nor can it turn; thus when signaling we can therefore communicate to great distances on one condition, only if we stand at high positions. With long waves, diffraction is sufficiently large to make it bypass the terrestrial convexity: we can therefore communicate with points that cannot be seen, and thus the main difficulty has disappeared, which has limited range capability. Hence, with optical telegraphy we can reach 40 km or 50 km, and by choosing favorable items (suitable machinery), we can reach, according to Poincaré, 300km.

---

240; p. 226; see also: Stachel, John, "Einstein's Clocks, Poincare's Maps/Empires of Time", *Studies in History and Philosophy of Science* B 36 (1), pp. 202-210.

## 11. The 1900 Congresses

At the turn of the century Poincaré had become one of France's most respected and honored mathematicians and scientists. His work in mathematical physics was well known and respected in scientific circles, and in addition the general public read his popular publications in science. In 1900 he was already member in fifteen foreign academies, Amsterdam, Berlin, Boston, Edinburgh, Stockholm, Copenhagen, San Petersburg, Rome, Munich, Washington, and the Royal Society of London.[132]

Poincaré was invited to give keynote lectures in congresses, and whenever people wanted to know the future direction of science they invited him as the sole authority to present a lecture on this topic. He had unshaken status as the oracle of science, and he won honor and respect from the best universities and received scientific prizes for his contributions.

In 1900, Paris hosted a grand event, *l'Exposition Universelle*. Paris was the center of art and science in Europe. On April 14 the universal exposition was inaugurated, a grandiose world fair that would last until November that year. Millions of visitors came to this exposition. Governments, industrial firms, entrepreneurs and individual scientists alike participated in the exposition and exhibited and promoted their goods, visions they could offer for other people's future, and scientific advances. Everything promised enterprise on a scale never before realized. Fair organizers budgeted for a projected 60 million visitors through seven months.[133]

As part of the world Fair, Paris organized that year 127 conferences dealing with topics such as History of Religion, the Second Mathematics International Conference (Poincaré participated in this conference, and was also in the organizing committee of this conference), the Thirteenth Medical International Conference, and the International Conference of Electricity.[134] Two additional international congresses that took place during the world fair, and are relevant to this discussion are: one congress dedicated to physics and the other to philosophy. In both congresses Poincaré was a central figure. The French Physical Society had begun preparations for the international congress of physics already in January 1899.[135] The same holds for the other congresses and for that of philosophy.

The first international congress of philosophy was held on August 1-5, 1900, under the presidency of Emile Boutroux, Poincaré's brother in law, who was at that time professor at the Sorbonne.[136]

A day after the congress of Philosophy ended, Poincaré hurried to the international congress of physics, taking place on August 6-12, 1900. The organizing committee consisted of Alfred Cornu (the professor from his student days at the Polytechnique), Louis-Paul Cailletet, Lucien Poincaré (his cousin) and Charles-Éduard Guillaume, and a host of prominent French physicists.[137]

Later this year on December 10, 1900, Poincaré traveled to Leiden, to the University of Leiden to Lorentz. There he met the most notable scholars of the time. The University celebrated the Festschrift on the occasion of the 25th anniversary of Lorentz's doctorate. Lorentz sat in the audience and the notable scholars talked one after the other on topics pertaining to his studies, notably his theory of electron. Poincaré chose to speak on the "Theory of [electron of] Lorentz and the Principle of Reaction".[138]

In 1900 Poincaré was indeed in his highest ranks, and he was the most successful scientist in France and maybe in the whole world. However, Poincaré felt deep inside a very big crisis. The contents of his lectures, which he presented in the two international conferences (of physics and philosophy), and the talk presented in the Lorentz Festschrift celebrations, reveal this crisis pertaining to reconciling Lorentz's theory with the principle of relativity and the principle of reaction.[Q]

## 12. Conventionalism and General Writings

Besides technical works, Poincaré regularly published papers in popular science and philosophy journals. He discussed the role of logic in mathematics, and the foundations of geometry and arithmetic; the foundations of mechanics, and the recent developments of physics.[139]

He developed the philosophy of conventionalism of geometry and the conventionalism of the principles of mechanics. In 1887 Poincaré published his first paper discussing the foundations of geometry, and in 1891 he published his second paper on the topic explicitly discussing conventionalism of geometry.[140]

Poincaré began to entertain conventionalist ideas in relation to geometry when he studied group theory in mathematics. While trying to solve the problems in group theory, Poincaré formulated for the first time *the philosophy of conventionalism for geometry* at the end of the paper, "Sur les hypothèses fondamentales de la géometrie" (The Fundamental Hypotheses of Geometry) of 1887 (without yet speaking of "conventionalism"),[141]

Poincaré assumed a group of movements not altering the distances, and arrived at a conclusion according to which geometry was nothing but the study of a group. Because the existence of one group is not incompatible with that of another group, the truth of the Euclidean geometry is not incompatible with that of the geometry of Lobachevsky, for example, and in fact with that of any other non-Euclidean geometry.

---

[Q] It is interesting to note that Einstein had a crisis at about the same time. Einstein appeared to have been trying to solve the conflict between the principle of Galilean relativity and that of the constancy of the velocity of light in Maxwell's theory; and the conflict between the principle of Galilean relativity and Maxwell's theory and Faraday's law. Although both Einstein and Poincaré were feeling a crisis at about the same time, they followed completely different routes.

We choose, out of all the possible groups, a particular group, with respect to which we relate the physical phenomena. This is the same as choosing, among the different coordinate systems, three coordinate axes with respect to which a geometrical figure is related. What determines this choice? It is first and foremost the simplicity of the chosen group. There is, however, another reason: there exists in nature remarkable bodies which we call *solids*, and experience has taught us that the diverse possible movements of these bodies are linked, to quite a great extent, by the same relations as the diverse operations of the chosen group. However, the chosen group is only more convenient than the others. One cannot speak of the Euclidean geometry as true and of the geometry of Lobachevsky as false; this is exactly the same as not being able to speak of the Cartesian coordinates as true and the Polar ones as false.

Poincaré proposed another geometry, the truth of which was not incompatible with the other geometries; he called it the "fourth geometry". The first time that Poincaré's fourth geometry appeared in print was in his 1891 paper, "Les Géométries non Euclidiennes".[142] Poincaré said that among the possible geometries the fourth geometry was one that deserves attention, because we can construct a fourth geometry in addition to those of Euclid, Lobachewski, and Riemann.

In an English paper, "On the Foundations of Geometry", published in *The Monist* in 1898 (an English translation by T. J. McCormack of a manuscript written by Poincaré), Poincaré explained for the first time Conventionalism using philosophical language,[143]

"Geometry and Contradiction.

In following up all the consequences of the different geometrical axioms, are we never led to contradictions? The axioms are not analytical judgments *a priori*; they are conventions. Is it certain that all these conventions are compatible?

These conventions, it is true, have all been suggested to us by experiments, but by crude experiments".

Poincaré summarized the paper and explained that geometry is not an experimental science, because it is in fact a study of a mathematical group,[144]

"experience forms merely the occasion for our reflecting upon the geometrical ideas which pre-exist in us. But the occasion is necessary; if it did not exist we should not reflect; and if our experiences were different, doubtless our reflections would also be different. Space is not a form of our sensibility; it is an instrument which serves us not to represent things to ourselves, but to reason upon things.

What we call geometry is nothing but the study of formal properties of a certain continuous group; so that we may say, space is a group".

In 1921 Einstein responded to Poincaré's conventionalism of geometry. Let us see how Einstein interpreted Poincaré. In his talk "Geometrie und Erfahrung" (Geometry and Experience), presented to the Prussian Academy of Sciences on January 27, 1921, Einstein started by implicitly presenting Kant's problem – the axioms of geometry are free creations of the human mind (Schöpfungen des menschlichen Geistes), and are independent of experience.[145] They thus do not refer to reality, and cannot say anything about reality.[146] Subsequently Einstein explained that geometry means "Geodesy", and geodesics are measurements, which are done with rigid bodies.[147] Einstein then added a theorem (Satz):[148]

"Solid bodies are related with respect to their possible dispositions, as are bodies in Euclidean Geometry of three dimensions; then the propositions of Euclidean Geometry contain statements about the behavior of practically rigid bodies".

Einstein then concluded that geometry thus supplemented is a natural-science, and, "We can virtually consider it as the oldest branch of physics".[149]

Poincaré wrote, experience has taught us that the diverse possible motions of solid bodies, and these bodies are linked, to quite a great extent, by the same relations as the diverse operations of the chosen group (and he chose Euclidean geometry). The chosen group is only more convenient than the others.[150]

Notice the difference between Einstein and Poincaré: Poincaré's starting point is group theory, while Einstein speaking about "praktische Geometrie" as opposed to "purely axiomatic Geometry" does not take this into account when responding to Poincaré's ideas.[151]

Let us ask the following question: what is the geometry of the world? For Poincaré this question has no meaning; we chose the geometry, because the chosen group is more convenient and simple than the others. But following Einstein's above Satz, within practical geometry this question has clear meaning.[152] And the punch line is, "To the conception of geometry that has been described, I attach particular importance for me, because without it, it would have been impossible for me to develop the theory of relativity".[153]

Einstein then referred to his rotating disk thought experiment from general relativity, without talking about a "disk": in a rotating system, which is rotating with respect to an inertial system, the laws of the rigid bodies do not correspond to the rules of Euclidean geometry because of the Lorentz-contraction.[154] Einstein used here the problematic concept of a rigid body in relativistic physics, while he already knew that there was no rigid body in relativistic physics.

Subsequently, Einstein debated explicitly with Henri Poincaré and his persistent adherence to Euclidean geometry; the pillar that held his geometric conventionalism: if we reject the relation between the body of axiomatic Euclidean geometry and the practically-rigid body of reality, we easily arrive at the following view, which should particularly be paid homage to the acute and profound thinker, H. Poincaré: Of all other conceivable axiomatic geometries, Euclidean Geometry is distinguished by its simplicity".[155]

Einstein did not agree with this standpoint because axiomatic geometry did not contain any statements about reality, and it could not be perceptible and testable, unless it was accompanied by physical theorems. Einstein said that, the problem posed by Poincaré was, if experience led to contradictions with the physical laws, then we could always change these and not Euclidean geometry, the simplest of all geometries. If geometry (G) does not say anything about the behavior of real things, but only geometry together with the epitome (P) of physical laws, then (G) + (P) is subject to the control of experience. Hence, we may choose (G) arbitrarily, and also parts of (P), because all these are conventions. We must choose (G) + (P) together to be in accord with experience.[156] This way we can always choose (the simplest geometry) Euclidean geometry and change (P).

Michel Paty commented on Einstein's presentation of Poincaré's standpoint, "Actually this is not exactly Poincaré's point of view, but a translation of it made by Einstein in his own perspective, that is according to his conception of physical Geometry. For, in Poincaré's conception, Geometry enters in the considerations of Physics only through definitions and is not on an equal footing with it".[157]

Paty is right. Einstein appeared not to fully comprehend Poincaré's geometric group theory, and thus he misrepresented his geometric conventionalism. Let us again call geometry (G), and the laws of optics (P) and weave these into Poincaré's following quotation from *Science and Hypothesis*,[158]

"We could renounce the Euclidean geometry [(G)] or better modify the laws of optics [(P)] and admit that light is not rigorously propagated in a straight line. It is needless to add that everybody would regard this solution as more advantageous. Euclidean geometry [(G)] has therefore nothing to fear from new experiments".

In the above quotation Poincaré suggested two possibilities: (1) if we renounce (G) then according to (P) light propagates in curved lines. (2) If we accept (G) then we modify (P). Both alternatives (1) and (2) are logically and empirically equivalent. It is *not* (G) + (P) one unit, but (G) *and* (P). Poincaré chose the second alternative (2), and he chose the group (G), because it is more convenient and simpler than the groups, non-Euclidean geometry.

Einstein then "sub specie," and, in essence, agreed with Poincaré (at least with "the Poincaré" that he had presented).[159] And he suggested an alternative view to this "Poincaré",[160]

The question whether the continuum is Euclidean or rather according to the general Riemannian scheme, or whether its structure is according to a different geometry, is actually a physical question, which must be answered by experience. This is not a question which is to be chosen by convention as a matter of convenience. Riemann's geometry is valid if the disposition laws of the practically rigid body are transformable into the laws of bodies of the Euclidean geometry, this becomes exact the more the dimensions of space-time in the considered area diminish. Einstein said that without this alternative view he would not have been able to formulate the theory of relativity.[161]

And Hans Reichenbach commented on the above text,[162]

"Those who know Einstein's clear and comprehensive way of thinking from personal conversations will understand the significance of this remark. Einstein is not a formal mathematician concerned with developing purely mathematical theories; rather, he thinks analytically, i.e., he is concerned with clarifying the meaning of concepts. Mathematics is, for him, only a means of expressing an intimate process – a process which operates from unconscious sources and for which the formal language is merely the framework".

At first indeed scientists were a little skeptical about Poincaré's conventionalism, and they were confused. This is clearly seen in Paul Painlevé's reaction to Poincaré's conventionalism of principles.

During the 1900 universal Exposition in the philosophy congress, Poincaré gave a talk, "Sur les principes de la mécanique", in the session devoted to mathematics-logic and philosophy and history of sciences. The audience of Poincaré's talk was mainly mathematicians. The session was directed by Professor Jules Tannery, Emil Boutroux's friend. The papers which were presented to this section were published a month later in September, 1900 in the *Revue de Métaphysique et de Morale* devoted to the congress of philosophy. Among the participants of the session were: Bertrand Russell, "The idea of order and absolute position in space and time", Giuseppe Peano who talked about "Mathematical definitions", Jacques Hadamard who spoke about "On induction in mathematics", and of course Poincaré who gave the above mentioned talk, "On the Principles of Mechanics".[163]

Poincaré read parts of his paper to the audience, and this was the first time that he had presented his *philosophy of conventionalism of principles*. One of the mathematicians sitting in the audience was the mathematics professor from Princeton by the name Edgar Odell Lovett (born in 1871 and died in 1957). A year later he reported what he had heard in this congress, and reported the discussions and responses to the talks of the speakers.[164] Lovett started by saying,

"In reading extracts from his memoir Professor POINCARÉ said that mechanics is an experimental science. But have not its principles an empirical and approximate truth only? That is the question. The principle of inertia is not an a priori truth; nor is it an experimental law, since we can never verify it. Similarly for the law of acceleration, which is simply the definition of force".[165]

Then Poincaré said, "The principle of reaction breaks up into an axiom (the uniform rectilinear motion of the center of gravity of an isolated system when constant coefficients are attributed to its elements) and a definition (that of mass), but we cannot verify the axiom in question because *we* are not in possession of an isolated system. It is approximately true for systems approximately isolated, but the question of knowing whether it is rigorously true for systems rigorously isolated is devoid of meaning".[166]

Poincaré, according to Lovett, went on now to treat the principle of relative motion,[167]

"The principle of relative motion seems to impose itself upon the mind and to be confirmed by experience; as a matter of fact, however, we can demonstrate it neither a priori nor a posteriori. M. Poincaré discussed in this connection Newton's argument in support of absolute motion. Finally the principle of the conservation of energy can be neither verified nor disproved by experience, since it reduces at bottom to this : 'There is something which remains constant' which is the very formula of determinism'".

According to Lovett, "M. Poincaré concludes that the principles of mechanics are from one point of view truths founded on experience and from another, a priori and universal postulates. In a word they are conventions, not absolutely arbitrary, but convenient, that is to say appropriate to experience. Thus is explained the fact that experience can construct or suggest the principles of mechanics, but can never overthrow them".[168]

Lovett's report of Poincaré's talk was succinct. Poincaré probably spoke the standard 20 minutes conference talk. He later published this talk, adapted it and expanded it somewhat, the result being different from Lovett's above report.[169]

According to Lovett, the discussion after Poincaré's talk began with a question by Painlevé,[170]

"who insisted upon the arbitrary character assumed by the principles of mechanics in M. Poincaré's exposition. They are conventions which experience can never bring to default: because as soon as any fact should contradict them we would always find, nolens volens, a means of adapting them to the new fact."

Painlevé gave an example for his claim,[171]

"For example, the law of gravitation is verified by a multitude of observations, but in other cases it appears at fault; we explain this divergence by saying that the bodies in question are electrified, or magnetic, etc., and we measure these new phenomena precisely by the discrepancy between the true attraction or repulsion and the Newtonian attraction."

However, Painlevé thought that Poincaré presented Newton's law only as convention,[172]

"It would seem then that the law of Newton is only a convention that the facts never contradict, because when they seem to contradict it, we invent new facts to justify it. Still, who would dream of replacing Newton's law by the following convention: 'Two bodies repel each other proportionally to their distance and inversely as their masses', correcting the divergence between this and experience by means of supplementary hypotheses?"

Painlevé reasoned, "We feel that the law of Newton is a convention preferable to all others, because it is clearly imposed by the facts." [173] According to Painlevé' this was exactly the case with Euclidean geometry: All the groups were equivalent, but it appeared that one was always preferable, because we could always change the auxiliary hypotheses so as to save it from fall. According to Poincaré's reasoning all groups were equivalent because we could just as well choose the non-Euclidean group and not change any auxiliary hypotheses. It was more convenient to choose the Euclidean group.

Lovett summarized Painlevé's standpoint, [174]

"To sum up, M. Painlevé conceives physical science as a method of successive approximations, oriented initially by empiricism and guided by certain principles of experimental origin. The "convergence" of this method is not assured a priori, but well justified by its success, i.e., by the more and more natural and perfect accord between theory and reality. In seeking the laws of nature, it is the divergence and increasing complications which give warning that we have lost our way.
M. Poincaré replied that there was really no lack of accord between M. Painlevé and himself. He himself recognized that science has always proceeded and will always proceed by successive approximations. But he was anxious to point out the series of artifices, more or less conscious, by which the founders of mechanics had succeeded in transforming the first approximation, not into a provisional truth susceptible of correction, but into a definitive and rigorous truth; and this to a great improvement in clearness of statement, and consequently to the benefit of science itself".

The next comment was made by Hadamard. He said that if "we assign as the object of mechanics, not the explanation of the phenomena of motion but merely their description in the simplest and most exact manner, the principles of this science, as we state them, are sufficiently justified". Hadamrad suggested that when we find facts in apparent contradiction with these principles we can suggest that a new force intervene, in place of changing the general principles.[175] In fact Poincaré's disk thought experiment is a good example of invoking such a force or mechanism, of which the inhabitants of the disk are unaware.

Pierre Duhem appeared to have objected to this suggestion when he subsequently remarked, "It is not a single determinate hypothesis, but the ensemble of the hypotheses of mechanics, that we can attempt to verify experimentally".[176]

Poincaré replied to Hadamard and Duhem,[177]

"I agreed that the experimental sciences can never verify anything but an ensemble of hypotheses. Every experiment furnishes us one equation in a very great number of unknowns. Our science, still imperfect, does not give us a sufficient number of equations; we have fewer than there are unknowns.

We can count on new experiments to give us continually new equations, which will diminish the indeterminacy of the problem. But as regards the unknowns introduced by geometry (curvature of space) or by rational mechanics (its most general principles), there is something more. Not only does experience not give us enough equations to determine them, but it is absurd and contradictory to suppose that it can ever give them; for the reason that these unknowns enter into the experimental problems as auxiliary supererogatory variables. This explains why the hypotheses which one might make relative to these unknowns are neither true nor false".

Mathematicians in the audience were not persuaded by Poincaré's answers. Another mathematician,[178]

"protested against M. Poincaré's skepticism. He held that the laws of mechanics have an objective value, and that they are not creations of the human intellect. The world existed before humanity, and the world will exist after it. It already obeyed, and it will continue to obey, the laws of mechanics. Hence science is true in the sense that it deals with real existences. M. Poincaré remarked that we raise here the question of the reality of the external world, which would be more in place in the first section (metaphysics)".

With this ended the first presentation of Poincaré's conventionalism of the principles of mechanics. Poincaré published an extended version of his talk from the philosophy congress later that year.[179] In 1902 Poincaré included this paper in his book *La Science et l'Hypothèse*.

Poincaré wrote a number of brilliant papers on the foundations of Geometry, group theory, and he also supplemented David Hilbert's work on the Foundations of Geometry. For his studies in this domain, in 1904, he was awarded the Lobatchevsky gold medal of the Mathematical Society of Kasan, and in 1905 the Bolyai Prize by the Hungarian Academy of Sciences.[180]

In 1902, the editor Camille Flammarion convinced Poincaré to collect and edit his different general writings and the published lectures.[181] Among the edited material one could find the papers on the foundations of geometry and mechanics, the papers after the lectures presented at the international congresses during the World Fair of 1900; the paper after the talk he delivered in the international congress of Arts and Science during the Saint Louis Louisiana celebration in 1904; and the 1898 paper he wrote during his work at the Bureau des Longitudes in Paris.

Poincaré took these papers and combined them into three books. He first created the best selling *La Science et l'Hypothèse*, from the papers that were published until 1902. He did little if any editing to the original papers; the only editing is simply organizing them into chapters of the book. The second book was *La Valeur de la Science*. This book was based on material published after 1902 and until 1905. And the third book was *Science et Méthode*, which was based on material published until 1908. All three books together were published as the series *Bibliothèque de Philosophie*. An additional volume was published posthumously in 1913, *Dernières Pensées*. The books were translated into many languages quite immediately after their appearance.[182]

As to the talk that Poincaré delivered in the international congress of Arts and Science during the Saint Louis celebration in 1904 (mentioned above). In 1904 there was an international exposition in Saint Louis, Missouri known by the name "The Saint Louis World Fair" or Louisiana Exposition. Towards the end of the 1904 summer, when Poincaré was 50 years old he went to the international Congress of Arts and Science held during this fair on September 19-25. There he gave a keynote lecture on September 24, a day before the congress was closed: on the present state and future of mathematical physics. Poincaré discussed the future of science in light of the deep crisis in mathematical physics. The crisis that Poincaré felt was so deep, "These principles, on which we have built everything, are they about to crumble down in their turn? Since some time, this may well be asked".[183]

## 13. The End

Delury wrote,[184]

"While at Rome [April], at the Mathematical Congress of 1908, he was taken seriously ill, and was unable to be present at the later sessions of the meeting. He was, in time, able to resume his work, and his many friends came to feel that, comparatively young, he had many years before him. This, however, was not to be. The ailment returning or persisting, a surgical operation was counseled. For a short time after the operation, all seemed well, but unexpectedly and suddenly the end came".

Poincaré died on July 17, 1912, just when he was so excited of the new quantum theory. He almost won the Nobel Prize. In 1910 Lorentz and Pieter Zeeman proposed Poincaré's name for the Nobel Prize in physics. However, Poincaré never won the Prize.[185]

The Royal Society of London sent to Poincaré's Funeral its representative Sir Joseph Larmor, whose theory of the electron Poincaré objected so fiercely.[186]

One of the undated condolences letters sent to Poincaré's cousin, Raymond Poincaré, was the following letter sent in French,[187]

"Prof. Dr. Albert Einstein

Member of the academy of sciences

I desire to present my homage to M. president of the Council on his cousin the grand master Henri Poincaré

                                        Albert Einstein"

Prof. Dr. Albert Einstein

Mitglied Der Akademie D. Wissenschaften

désirerait présenter les hommages à M. Le président du Conseil et au cousin de notre grand maître Henri Poincaré.

A. E.

I wish to thank Prof. John Stachel from the Center for Einstein Studies in Boston University for sitting with me for many hours discussing special and general relativity and their history. Almost every day, John came with notes on my draft manuscripts, directed me to books in his Einstein collection, and gave me copies of his papers on Einstein, which I read with great interest

# Endnotes

[11] Volterrra, Vito, *Henri Poincaré*, 1914, Paris: F. Alcan.

[12] Appell, Paul (1912), "Henri Poincaré, en mathematiques specials a Nancy", *Acta Mathematica* 38, 1921, pp. 189-195 (written on December 22, 1912 shortly after Poincaré's death).

[13] Appell, Paul, *Henri Poincaré*, 1925, Paris: Plon.

[14] Dantzig, Tobias, *Henri Poincaré Critic of Crisis Reflections on his Universe of Discourse*, 1954, New York and London: Charles Scribner's Sons.

[15] Bellivier, *Heari Poincare ou la vocation souveraine*, 1956, Paris: Gallimard.

[16] Bottazzini, Umberto, "Adolescence à Nancy", *Poincaré Philosophe et mathematician, Les Génies de la Science*, *Pour la Science*, *Scientific American, edition Française*, Novembre 2000, pp. 32-37.

[17] Bottazzini, 2000, pp. 32-37.

[17] Walter, Scott, Etienne Bolmont et André Coret (eds), *La correspondance entre Henri Poincaré et les physiciens, chimistes et ingénieurs*, Birkhäuser: Basel, 2007, Walter, Scott, "Henri Poincaré's Student Notebooks, 1870-1878", *Philosophia Scientiae* 1, 1996, pp. 1-17; Miller, Arthur I. *Insights of Genius: Imagery and Creativity in Science and Art*, 1996, New York: Springer; 2000, Cambridge, Mass: MIT Press, p. 341.

[18] Mawhin, Jean, "Henri Poincaré: a life at the Service of Science", *Proceedings of the Symposium Henri Poincaré*, Brussels, 8-9, October, 2004, pp. 1-17.

[19] Galison, Peter, *Einstein's Clocks, Poincaré's Maps Empire of Time*, 2003, New York: W. W. Norton & Co.

[20] Toulouse, 1909, pp. 35-36.

[21] Toulouse, 1909, p. 124.

[22] Toulouse, 1909, p. 140.

[23] Toulouse, 1909, p. 142.

[24] Dantzig, 1954, p. 2.

[25] Toulouse, 1909, p. 142.

[26] Toulouse, 1909, p. 142.

[27] Mawhin, Jean, "Henri Poincaré: a life at the Service of Science", *Proceedings of the Symposium Henri Poincaré*, Brussels, 8-9, October, 2004, pp. 1-17, p. 2.

[28] Lebon, 1912, p. 6. "De votre nom—Pontcaré, plutôt que Poincaré, car avez-vous dit,".

[29] Mawhin, 2004, p. 2; "Death of Henri Poincaré. Mathematician and Philosopher", *The Times*, 17 July, 1912; Toulouse, pp. 15-16.

[30] Toulouse, 1909, p. 158. "il y a lieu de noter que plusieurs des sujets mâles de la famille ont été des individus distingués".

[31] Toulouse, 1909, p. 17.

[32] Toulouse, 1909, p. 21. "M. H. Poincaré aurait commencé à parler à neuf mois".

[33] Lebon, 1912. P. 7; Delury, p. 310. And the original version, "Vous n'avez guère attendu pour révéler votre vocation et l'on vous citera justement comme le plus précoce des enfants prodiges. Vous aviez neuf mois, lorsque, pour la première fois, la nuit venant, vos yeux se portèrent sur le ciel. Vous y avez

38vu s'allumer une étoile. A votre mère, qui était aussi votre nourrice, vous avez montré avec obstination ce point qui brillait. Vous en avez découvert un deuxième, et ce fut le même étonnement et ce cri de votre raison: *Encore là-bas!* Au troisième, au quatrième, pareil cri de joie et pareil enthousiasme; il fallut vous coucher, tant vous vous excitez à chercher des étoiles. Ce soir-là, vous aviez pris votre premier contact avec l'infini et vous aviez inauguré vos cours d'astronomie: on ne saurait professer plus jeune".

[34] Lebon, 1912, pp. 7-8.

[35] Delury, p. 310.

[36] Toulouse, 1909, p. 21.

[37] Toulouse, 1909, p. 22.

[38] Toulouse, 1909, p. 23.

[39] Lebon, 1912, p. 8; Delury, p. 310; Toulouse, 1909, p. 23.

[40] Toulouse, 1909, p. 23.

[41] Delury, p. 310.

[42] Lebon, 1912, p. 8.

[43] Toulouse, 1909, p. 140.

[44] Toulouse, 1909, p. 24.

[45] Lebon, 1912, p. 8; Delury, p. 310.

[46] Delury, p. 311.

[47] Toulouse, 1909, p. 28; Lebon, 1912, p. 8.

[48] Toulouse, 1909, p. 26.

[49] Dieudonné, Jean, "Poincaré", *Dictionary of Scientific Biography*, 1970-1990, New-York: Scribner.

[50] Toulouse, 1909, pp. 24-25. Il a appris l'allemand au lycée. Dans sa famille, on ne parlait pas allemand. Il s'est perfectionné dans cette langue pendant la guerre et aussi au cours d'un séjour de trois mois qu'il fit en Autriche en 1877.

[51] Delury, p. 311.

[52] Lebon, 1912, pp. 9-10.

[53] Dantzig, p. 4.

[54] Poincaré, Henri, "La mécanique nouvelle" (Last Wolfskehl lecture held at Göttingen, 1909), Sechs vorträge über ausgewählte gegenstände aus der reinen mathematik und der mathematischen physik, auf Einladung der Wolfskehl-Kommission der Königlichen Gesellschaft der Wissenschaften gehalten zu Göttingen vom 22-28. April 1909 von Henri Poincaré, Mitglied der Französischen Akademie Professor an der Faculté des Sciences der Universität Paris, 1910, Leipzig: Teubner, pp. 41-47.

[55] Toulouse, 1909, p. 25.

[56] "J'ai demandé à M. H. Poincaré quelles étaient ses opinions sur les questions les plus Courantes. Au point de vue religieux, il croyait au moment de sa première communion, puis, progressivement, le doute est venu et, vers l'âge de dix-huit ans, il a cessé de croire. Il est pour la libre pensée, pour le droit de rechercher et de dire la vérité, et, pour cela, opposé à l'intolérance cléricale". Toulouse, 1909, p. 143.

---

[57] Appell, 1912, p.190.

[58] Appell, 1912, p.189.

[59] "voilà un type très fort, il vient d'être reçu second à l'École Forestière".

[60] "Vous l'emportiez en mathématiques sur tous vos concurrents de Paris et des départements; il ne tenait qu'à vous d'entrer, le deuxième de la promotion, à l'École forestière, autre gloire de Nancy". Lebon, 1912, p. 10.

[61] Appell, 1912, p.190.

[62] Appell, 1912, p.191.

[63] Appell, 1912, p.191.

[64] Appell, 1912, p.193.

[65] Appell, 1912, p.193.

[66] Appell, 1912, pp.193-194.

[67] Appell, 1912, p. 193.

[68] Appell, 1912, p.194.

[69] Appell, 1912, p.194.

[70] Appell, 1912, p.194.

[71] Appell, 1912, pp.194-195.

[72] Bottazzini, 2000, pp. 35-36.

[73] Galison, 2003, p. 50.

[74] "A l'École Polytechnique, M. H. Poincaré suivait — paraît-il — les cours de mathématiques sans prendre de notes". Toulouse, 1909, p. 99.

[75] "On raconte que vous avez suivi vos cours, au moins de mathématiques, sans prendre une note, sans regarder, ni même recueillir les feuilles autographiées qui reproduisent l'exposé du professeur". Lebon, 1912, p. 10.

[76] Walter, 1996, p. 4.

[77] Bottazzini, 2000, p. 36.

[78] Walter, 1996, p. 6.

[79] Lebon, 1912, p. 11.

[80] Walter, 1996, p. 4.

[81] "Il fut reçu le premier à l'École polytechnique. Il en sortit avec le n° 2.". Toulouse, 1909, p. 28.

[82] Toulouse, 1909, p. 28.

[83] Walter, 1996, p. 4.

[84] Toulouse, 1909, p. 142.

[85] Walter, 1996, p. 4.

[86] Based on a memorial read by the president of the Academy of Sciences, M. Troost in the Academy of Sciences on May 5, 1905, and published in the *Annales des Mines* 7, 1905.

[87] Walter, 1996, p. 4.

[88] Toulouse, 1909, p. 27.

[89] "Ce n'est point que, connaissant vos distractions, votre mère vous vît partir sans inquietude". Lebon, 1912, p. 11.

[90] Lebon, 1912, p. 11.

[91] Mawhin, 2004, p. 3.

[92] Walter, 1996, p. 4.

[93] Mawhin, 2004, p. 3.

[94] Miller, 1996, p. 343.

[95] Mawhin, 2004, pp. 4-5.

[96] "Depuis quinze jours je m'efforçais de démontrer qu'il ne pouvait exister aucune fonction analogue à ce que j'ai appelé depuis les fonctions fuchsiennes; j'étais alors fort ignorant; tous les jours, je m'asseyais à ma table de travail, j'y passais une heure ou deux, j'essayais un grand nombre de combinaisons et je n'arrivais à aucun résultat. Un soir, je pris du café noir, contrairement à mon habitude, je ne pus m'endormir; les idées surgissaient en foule; je les sentais comme se heurter, jusqu'à ce que deux d'entre elles s'accroch assent pour ainsi dire pour former une combinaison stable. Le matin, j'avais établi l'existence d'une classe de fonctions fuchsiennes, celles qui dérivent de la série hypergéométrique ; je n'eus plus qu'à rédiger les résultats, ce qui ne me prit que quelques heures.

Je voulus ensuite représenter ces fonctions par le quotient de deux séries; cette idée fut parfaitement consciente et réfléchie; l'analogie avec les fonctions elliptiques me guidait. Je me demandai quelles devaient être les propriétés de ces séries si elles existaient, et j'arrivai sans difficulté à former les séries que j'ai appelées thétafuchsicnnes.

A ce moment, je quittai Caen, que j'habitais alors, pour prendre part a une course géologique entreprise par l'École des Mines. Les péripéties du voyage me firent oublier mes travaux mathématiques ; arrivés à Coutanccs, nous montâmes dans un omnibus pour je ne sais quelle promenade; au moment où je mettais le pied sur le marche-pied, l'idée me vint, sans que rien dans mes pensées antérieures parût m'y avoir préparé, que les transformations dont j'avais fait usage pour définir les fonctions fuchsiennes étaient identiques à celles de la géométrie non euclidienne. Je ne fis pas la vérification; je n'en aurais pas eu le temps puisque, à peine assis dans l'omnibus, je repris la conversation commencée, mais j'eus tout de suite une entière certitude. De retour à Caen, je vérifiai le résultat à tête reposée pour l'acquit de ma conscience.

Je me mis alors à étudier des questions d'arithmétique sans grand résultat apparent et sans soupçonner que cela pût avoir le moindre rapport avec mes recherches antérieures. Dégoûté de mon insuccès, j'allai passer quelques jours au bord de la mer, et je pensai; a tout autre chose. Un jour, en me promenant sur une falaise, l'idée me vint, toujours avec les mêmes caractères de brièveté, de soudaineté et de certitude immédiate, que les transformations arithmétiques des formes quadratiques ternaires
indéfinies étaient identiques à celles de la géométrie non euclidienne".

Poincaré, 1908a, pp. 7-8.

[97] Hadamard, 1945, pp. 11-12.

[98] Hadamard, 1945, p. 15.

[99] "C'est là une méthode de travail peu commune en matière scientifique et elle constitue un caractère bien particulier de l'activité mentale de M. H Poincaré". Toulouse, 1909, p. 187.

[100] "Ce qui vous frappera tout d'abord ce sont ces apparences d'illumination subite, signes manifestes d'un long travail inconscient antérieur ; le rôle de ce travail inconscient dans l'invention mathématique me paraît incontestable, et on en trouverait des traces dans d'autres cas où il est moins evident Souvent quand on travaille une question difficile, on ne fait lien de bon la première fois qu'on se met à la besogne ; ensuite on prend un repos plus ou moins long, et on s'assoit de nouveau devant sa table". Poincaré, 1908a, p. 9.

[101] Poincaré, 1908a, pp. 7-8.

[102] Toulouse, 1909, pp. 144-145.

[103] "C'est là une méthode de travail peu commune en matière scientifique et elle constitue un caractère bien particulier de l'activité mentale de M. H Poincaré. De tous ces faits, il ressort que l'activité intellectuelle de M. H. Poincaré est surtout spontanée, automatique". Toulouse, 1909, p. 187.

[104] "nous le verrons plus loin – M. H. Poincaré n'est pas visuel". Toulouse, 1909, p. 14.

[105] Toulouse, 1909, pp. 101; 103.

[106] "M. H. Poincaré un procédé de recherche. Ainsi il ne fait pas de plan quand il écrit un mémoire". Toulouse, 1909, p. 186.

[107] Toulouse, 1909, p. 145.

[108] Toulouse, 1909, p. 27.

[109] Toulouse, 1909, pp. 145-146.

[110] Delury, p. 313-314.

[111] Poincaré, Henri, *Lecons sur la théorie mathématique de la lumière, professées pendant le premier semestre 1887-1888*, Cours de physique mathématique, edited by Jules Blondin,Cours de la Faculté des sciences de Paris, 1889, Paris: Carré et Naud; Poincaré, Henri, *Electricité et optique I. Les théories de Maxwell et la théorie électromagnétique de la lumière. Lecons professées pendant le second semestre 1888-1889*, Cours de la Faculté des sciences de Paris, Cours de physique mathématique, edited by Jules Blondin,1890, Paris: Carré et Naud; Poincaré, Henri, *Electricité et optique, II. Les théories de Helmholtz et les expériences de Hertz. Lecons professées pendant le second semestre 1889-1890*, Cours de la Faculté des sciences de Paris, Cours de physique mathématique edited by Bernard Brunhes, 1891, Paris: Carré et Naud; Poincaré, Henri, *Théorie mathématique de la lumière,II. Nouvelles études sur la diffraction. Lecons professées pendant le premier semestre 1891-1892*, Cours de la Faculté des sciences de Paris, Cours de physique mathématique, edited by M. Lamotte and D. Hurmuzescu, 1892 Paris: Carré et Naud; Poincaré, Henri, *Les oscillations électriques. Lecons professées pendant le premier trimestre 1892-1893*, Cours de physique mathématique, edited by Charles Maurain. Cours de la Faculté des sciences de Paris, 1894, Paris: Carré et Naud; Poincaré, Henri, *La théorie de Maxwell et les oscillations Hertziennes*, 1894, Paris: Carré et Naud; Poincaré, Henri, *Electricité et optique. La lumière et les théories électrodynamiques.Lecons professées la Sorbonne en 1888, 1890 et 1899,* cours de physique mathématique, edited by Jules Blondin and Eugène Néculcéa, 1901, Paris: Carré et Naud; deuxième edition, Paris: Gauthier-Villars.

[112] Poincaré, Henri, "Contribution à la théorie des expériences de M. Hertz", *Comptes rendus hebdomadaires de l'Académie des sciences* 111, 1890, pp. 322-326; Poincaré, Henri, "Contribution à la théorie des expériences de M. Hertz", *Archives des sciences physiques et naturelles* 24, 1890, pp. 285-288; Poincaré, Henri, "Sur les équations aux dérivées partielles de la physique mathématique", *American Journal of Mathematics* 12 , 1890, pp. 211-294; Poincaré, Henri, "Sur la résonance multiple des oscillations hertziennes", *Archives des sciences physiques et naturelles* 25, 1891, pp. 609-627; Poincaré, Henri, "Sur la théorie des oscillations hertziennes", *Comptes rendus hebdomadaires de l'Académie des sciences* 113, 1891, pp. 515-519; Poincaré, Henri, "Sur le calcul de la période des

excitateurs hertziens", *Archives des sciences physiques et naturelles* 25, 1891, pp. 5-25; Poincaré, Henri, "Sur la propagation des oscillations électriques", *Comptes rendus hebdomadaires de l'Académie des sciences* 114, 1892, pp. 1229-1233; Poincaré, Henri, "Sur la propagation des oscillations hertziennes. *Comptes rendus hebdomadaires de l'Académie des sciences* 114, 1892, pp. 1046-1048; Poincaré, Henri, "Sur la propagation de l'électricité", *Comptes rendus hebdomadaires de l'Académie des sciences* 117, 1893, pp. 1027-1032; Poincaré, Henri, "Les idées de Hertz sur la mécanique", *Revue générale des sciences pures et appliqués* 8, 1897, pp. 734-743; Poincaré, Henri, "Sur les excitateurs et résonateurs hertziens (à propos d'un article de M. Johnson)", *Éclairage électrique* 29, 1901, pp. 305-307.

[113] Poincaré to Hertz, August 15, 1890, letter 30.1, Walter, Bolmont, Coret, and Basel (ed), 2007, pp. 184-185.

[114] Hertz to Poincaré, August 21, 1890, letter 30.2, Walter, Bolmont, Coret, and Basel (ed), 2007, p. 186.

[115] Hertz to Poincaré, August 21, 1890, letter 30.2, Walter, Bolmont, Coret, and Basel (ed), 2007, p. 186.

[116] Poincaré to Hertz, September 11, 1890, letter 30.3, Walter, Bolmont, Coret, and Basel (ed), 2007, p. 187.

[117] Hertz to Poincaré, September 22, 1890, letter 30.4, Walter, Bolmont, Coret, and Basel (ed), 2007, p. 188.

[118] Hertz to Poincaré, September 22, 1890, letter 30.4, Walter, Bolmont, Coret, and Basel (ed), 2007, p. 189.

[119] Poincaré to Hertz, October 8, 1890, letter 30.5, Walter, Bolmont, Coret, and Basel (ed), 2007, p. 189.

[120] Hertz to Poincaré, October 19, 1890, letter 30.6, Walter, Bolmont, Coret, and Basel (ed), 2007, pp. 192-195.

[121] Hertz to Poincaré, December 30, 1891, letter 30.15, Walter, Bolmont, Coret, and Basel (ed), 2007, p. 202.

[122] Broglie de, Louis, "Henri Poincaré et les théories de la physique", *Savants et Découvertes*, 1951, Paris: Albin Michel, p. 45.

[123] Delury, p. 311.

[124] Galison, 2003, p. 129.

[125] Galison, 2003, p. 129.

[126] Galison, 2003, p. 221.

[127] "De Paris à Berlin, A envoie un signal télégraphique, avec un sans-fil, si vous voulez, pour être tout à fait modern". Poincaré, 1909, p. 44.

[128] Poincaré, Henri, *La théorie de Maxwell et les oscillations hertziennes; La télégraphie sans fil*, 1907, Paris: Gauthier Villars, third edition.

[129] Poincaré, Henri (1908b), "Conférences sur la Télégraphie Sans Fil", *La Lumière Électrique* 4, 1908, pp. 259-266, pp. 291-297, pp. 323-327, pp. 355-359, pp. 387-393 *La Lumiére Électrique* 4, Éditorial.

[130] Poincaré, Henri, "Sur Diverse Questions Relatives A la Télégraphie Sans Fil", *La Lumière Électrique* 13, 1911, pp. 7-12, pp. 35-40, pp. 67-72, pp. 99-104.

[131] Poincaré, 1907, pp. 81, 88-89.

[132] "Death of Henri Poincaré. Mathematician and Philosopher", *The Times*, 17 July, 1912.

[133] Staley, Richard, *Einstein's Generation. The Origins of the Relativity Revolution*, 2008, Chicago: The University of Chicago Press, p. 133.

[134] Staley, 2008, pp. 167-168.

[135] Staley, 2008, p. 166.

[136] Lovett, E. O., "Mathematics at the International Congress of Philosophy, Paris 1900", *Bulletin of the American Mathematical Society 7*, 1901, N. 4, pp. 157-183; p. 157.

[137] Staley, 2008, p. 166.

[138] Poincaré, Henri (1900b), "La théorie de Lorentz et le principe de réaction", *Archives néerlandaises des sciences exactes et naturelles. Recueil de travaux offerts par les auteurs à H.A.Lorentz* (The Hague: Nijhoff) Ser II (1900), 5, pp. 252-278.

[139] Mawhin, 2004, p. 9.

[140] Poincaré, Henri, "Sur les hypothèses fondamentales de la géometrie", *Bulletin de la société mathèmatique de France* 15, 1887, pp. 203-216; Poincaré, Henri, "Les géométries non-Euclidiennes", *Revue génerale des sciences pures et Appliqués* 2, 1891, pp. 769-774; translated to English by W.J.L, "Non Euclidean Geometry, *Nature*, Feb 25, 1892, pp. 404-407.

[141] Poincaré, 1887, in *oeuvres*, pp. 90-91.

[142] Poincaré, 1891, p. 772.

[143] Poincaré, Henri, "On the Foundations of Geometry", *The Monist* 9, 1898, pp. 1-43; translated from Poincaré's manuscript by T.J. McCormack, p. 38.

[144] Poincaré, 1898, p. 41.

[145] Einstein, Albert, "Geometrie und Erfahrung, *Preußische Akademie der Wissenschaften* (Berlin). *Sitzungsberichte* I, 1921, pp.123-130; p. 124.

[146] Einstein, 1921, p. 125.

[147] Einstein, 1921, p. 125.

[148] Einstein, 1921, p. 125.

[149] Einstein, 1921, p. 125.

[150] Poincaré, 1887, in *oeuvres*, pp. 90-91.

[151] Einstein, 1921, p. 125.

[152] Einstein, 1921, p. 125.

[153] Einstein, 1921, p. 126.

[154] Einstein, 1921, p. 126.

[155] Einstein, 1921, p. 126.

[156] Einstein, 1921, pp. 126-127.

[157] Paty, Michel, "Physical Geometry and Special Relativity: Einstein and Poincaré ", in L. Boi, D. Flament et Salanski, J.M. (eds.), 1830-1930: un siècle de géométrie, de C.F. Gauss et B. Riemann à H.

Poincaré et E. Cartan. Epistémologie, histoire et mathématiques, Springer-Verlag, Berlin, 1992, p. 126-149; p. 132.

[158] Poincaré, Henri (1902), *La science et l'hypothése*, 1902/1968, Paris: Flammarion (The second edition of 1968 is based on the one edited by Gustave Le Bon,1917; translated to English by W.J.G., *Science and Hypothesis* New York: The Walter Scott Publishing co, Feb 1905, New-York: Dover;1952, pp. 95-96.

[159] Einstein, 1921, p. 127.

[160] Einstein, 1921, p. 128.

[161] Einstein, 1921, p. 126; Einstein spoke about "the practically rigid body", even though he knew it had no meaning in the theory of relativity, and that there was no rigid body in practice.

[162] Reichenbach, Hans, "The Present State of the Discussion on Relativity", 1922 in Reichenbach, Hans, *Selected Writings, 1909-1953*, edited by Maria Reichenbach and Robert S. Cohen, Vol. Two, 1978, translated by Elizabeth Hughes Schneewind, Holland: D Reidel Publishing Company, p. 33.

[163] Lovett, 1901, pp. 157-158.

[164] Lovett, 1901.

[165] Lovett, 1901, p. 161.

[166] Lovett, 1901, p. 161.

[167] Lovett, 1901, p. 161.

[168] Lovett, 1901, p. 161.

[169] Poincaré, Henri (1900c), "Sur les principes de la mécanique (lecture before the international congress of philosophy in Paris, 1st, August 1900), *Bibliothèque du Congrèss international de philosophie* tenu à Paris du 1$^{er}$ au 5 aout 1900,Vol. III: *Logique et histoire des sciences*, 1901, Paris: Colin, pp. ,457-494; adapted in Poincaré, 1902, chapters. 6 and 7.

[170] Lovett, 1901, pp. 161-162.

[171] Lovett, 1901, p. 162.

[172] Lovett, 1901, p. 162.

[173] Lovett, 1901, p. 162.

[174] Lovett, 1901, pp. 162-163.

[175] Lovett, 1901, p. 163.

[176] Lovett, 1901, p. 163.

[177] Lovett, 1901, pp. 163-164.

[178] Lovett, 1901, pp. 164-165.

[179] Poincaré, 1900b.

[180] Delury, p. 318.

[181] Mawhin, 2004, p. 9.

[182] Mawhin, 2004, p. 9.

[183] Poincaré, Henri, "L'état actuel et l'avenir de la Physique mathématique", *Bulletin des sciences mathématiques*, 28, Décembre 1904, pp. 302-324; adapted in Poincaré, Henri, *La valeur de la science*, 1905/1970 Paris: Flammarion; translated to English by F. Maitland, *The Value of Science*, New-York: Dover, chapters.7,8 and 9; translated to English by G.B. Halsted: "The Principles of Mathematical Physics", *The Monist* 15, 1905, p. 1; 1904, p. 307.

[184] Delury, pp. 319-320; Poincaré, "Compte rendu d'ensemble des travaux du IVe Congrès des Mathématiciens tenu à Rome en 1908 *Le Temps* 48, pp. 2–3.

[185] Walter, Bolmont and Coret (ed), 2000, p. 260.

[186] Poincaré said in 1900: "However, Lorentz had no other ambition but to include in a single whole all the optics and electrodynamics of moving bodies; he made no pretensions to giving a mechanical explanation. Larmor goes farther; preserving the essentials of Lorentz's theory, he grafts MacCullagh's ideas on the direction of the movement of the ether. However ingenious this attempt may be, the fault in Lorentz's theory subsists, and is even aggravated. With Lorentz, we do not know what the movements of the ether are: thanks to this ignorance, we might suppose them such as compensating those of matter, […]. With Larmor, we know the movements of the ether and we can certify that the compensation does not take place. If Larmor has to my mind failed, does that mean that a mechanical explanation is impossible? Far from it: I said further above that as long as a phenomenon obeys the two principles of energy and of least action, it involves an infinite number of mechanical explanations; it is the same for optical and electrical phenomena". Poincaré, Henri (1900a), "Sur les relations entre la physique expérimentale et la physique mathématique", *Revue génerale des sciences pures et appliqués* 21, 1900, pp. 1163-1175; reprinted in Poincaré, 1902, chapters. 9 and 10; translated to English by G.K. Burges: "Relations between Experimental Physics and Mathematical Physics", *The Monist* 12, 1902, p. 516; 1900, p. 1173.

[187] Einstein, Albert, *Oeuvres Choisies, Correspondances Françaises*, Tome 4, Éditions Du Seuil, 1989, p. 256.